\documentclass[APS,PRB,twocolumn,floatfix]{revtex4}
\PassOptionsToPackage{linktocpage}{hyperref}
\usepackage{hyperref}
\usepackage{bm}
\usepackage{amssymb}
\usepackage{amsfonts}
\usepackage{amsmath}
\usepackage{graphicx}
\usepackage{dcolumn}
\usepackage{color}
\usepackage{setspace}
\usepackage{longtable}

\begin{document}
\title{Review of the Structural Stability, Electronic and Magnetic Properties of Nonmetal-Doped TiO$_2$ from First-Principles Calculations}

\author{Kesong Yang}
\email{kesong.yang@gmail.com}
\affiliation{School of Physics, State Key Laboratory of Crystal Materials, Shandong University, Jinan 250100, China}
\altaffiliation{Present Address: Duke University, NC 27708}

\author{Ying Dai}
\email{daiy60@sina.com}
\affiliation{School of Physics, State Key Laboratory of Crystal Materials, Shandong University, Jinan 250100, China}
\author{Baibiao Huang}
\affiliation{School of Physics, State Key Laboratory of Crystal Materials, Shandong University, Jinan 250100, China}

\begin{abstract}
This paper reviews and summarizes the recent first-principles theoretical
studies of the structural stability, electronic structure, optical and magnetic
properties of nonmetal-doped TiO$_2$. The
first section presents a comparison study of the structural stability for
X-anion and X-cation doped TiO$_2$ (X=B, C, Si, Ge, N, P, As, Sb, S, Se, Te, F, Cl, Br, and I), which reveals that
the sites of nonmetal dopants (i.e., at O sites or at Ti sites) in TiO$_2$ are determined by the growth
condition of doped TiO$_2$ and the dopants' electronegativities. The
next section reviews the electronic structure, optical absorption and mechanism
of the visible-light photocatalytic activity for nonmetal-doped TiO$_2$. The third
section summarizes the origin of the spin-polarization and the magnetic coupling
character in C- (N- and B-) doped TiO$_2$.
\end{abstract}

\maketitle

\tableofcontents

\section{Introduction}
Titanium dioxide (TiO$_2$) is one ideal semiconductor photocatalyst because of
its excellent properties (e.g.,  high activity, good stability, nontoxicity and
        low cost), which has many promising applications in the fields of
renewable energy and environmental protections.\cite{Fujishima_1972_Nature,
    Linsebigler_1995_Chem.Rev, Hoffmann_1995_Chem.Rev, Chen_2007_Chem.Rev}
    However, the larger band gap of TiO$_2$ ($\sim$3.2 eV for anatase and
            $\sim$3.0 eV for rutile) makes it inefficient for visible-light to
    excite the electron-hole pairs, which are necessary to initiate a photocatalytic process.\cite{Linsebigler_1995_Chem.Rev, Hoffmann_1995_Chem.Rev, Chen_2007_Chem.Rev} Hence, the photocatalytic applications of TiO$_2$ in the visible-light range are heavily limited. To improve the spectra response and photocatalytic activity of TiO$_2$ in the whole solar spectra, numerous efforts have been carried out and some progress is achieved.\cite{Linsebigler_1995_Chem.Rev, Hoffmann_1995_Chem.Rev, Chen_2007_Chem.Rev, Nie_2009_ijp_review, Zaleska_2008_RPOE, Henderson_2011_SSR} Especially in recent years, a number of attempts have been made to improve the visible-light absorption of TiO$_2$ by nonmetal doping, which either introduces some impurity states in the band gap or modifies the fundamental band gap of TiO$_2$, and promote the photocatalytic activity of TiO$_2$ to some degree.\cite{Asahi_2001_Science,  Khan_2002_Science, Chen_2007_Chem.Rev, Nie_2009_ijp_review, Zaleska_2008_RPOE, Henderson_2011_SSR}

In this paper, we review and summarize the recent theoretical progress on the structural
stability, electronic structures, optical and magnetic properties of
nonmetal-doped TiO$_2$. This review is divided into three sections. Firstly, we
illustrate the relationship between the doping sites of nonmetal atoms and their
electronegativities as well as the growth condition of doped TiO$_2$ samples.
Secondly, we focus on the nonmetal doping effects on the electronic structure
and optical absorption of TiO$_2$ as well as the modification of its
photocatalytic activity. Finally, we review the origin of the spin-polarization and magnetic coupling character in C- (N- and B-) doped TiO$_2$. This work aims to provide a general understanding on the structural stability, electronic structure, optical absorption and magnetic properties of nonmetal-doped TiO$_2$.

\section{Structural Stability}
First-principles density functional theory (DFT) calculations for the relative
stability of nonmetal-doped TiO$_2$ can help us understand the formation of the
doped structures and provide useful guidance to prepare
samples.\cite{VandeWalle_2004_JAP, DiValentin_2005_CM, Livraghi_2006_JACS,
    Yang_2006_JPCB, DiValentin_2007_CP,Yang_2007_JPCC_N, Yang_2007_JPCC_S,
    Yang_2008_CPL, Yang_2008_CM} Firstly, let us focus on the structural
    stability of substitutional doping models. In principle, there are two
    possible substitutional doping ways for X-doped TiO$_2$, i.e., an X anion at an
    O site (X@O) or an X cation at a Ti site (X@Ti). The defect formation energy required for X substituting for either O or Ti in TiO$_2$ could be calculated from the following formulas, respectively.\cite{Yang_2007_JPCC_S, Yang_2008_CPL, Yang_2008_CM}
\begin{equation}
E_f^X = {E_{X - doped}} - {E_{undoped}} - {\mu _X} + {\mu _O}
\end{equation}
\begin{equation}
E_f^X = {E_{X - doped}} - {E_{undoped}} - {\mu _X} + {\mu _{Ti}}
\end{equation}

$E_{X - doped}$ is the total energy of X-doped TiO$_2$ and $E_{undoped}$ is the
total energy of undoped TiO$_2$. ${\mu _X}$ is the chemical potential of dopant
X, and ${\mu _O}$ (${\mu _{Ti}}$) is the chemical potential of the O (Ti). The
chemical potentials of Ti and O depend on whether TiO$_2$ is grown under O-rich
or Ti-rich growth condition. Under Ti-rich condition, the Ti chemical potential
can be assumed as the energy of bulk Ti while the O chemical potential can be obtained by the growth condition:
\begin{equation}
{\mu _{Ti}} + 2{\mu _{O}} = {\mu _{Ti{O_2}}}
\end{equation}
Under O-rich condition, the chemical potential of O can be calculated from the ground-state energy of O$_2$ molecule, while the chemical potential of Ti is then fixed by condition (3). Therefore, a link between the defect formation energy and the external growth condition of doped TiO$_2$ can be created.

The relationship between the formation energies of substitutional nonmetal-doped
TiO$_2$ and its growth condition has been studied
systemically.\cite{DiValentin_2005_CM, Yang_2007_JPCC_N, Yang_2007_JPCC_S,
    Yang_2008_CPL, Yang_2008_CM} For Si-doped TiO$_2$, the formation energy of
    substitutional Si-cation doped model is much less than that of
    substitutional Si-anion doped model under both Ti-rich and O-rich growth
    conditions.\cite{Yang_2008_CPL, Long_2009_PCCP, Shi_2011_JSSC} This
    indicates that Si is energetically more favorable to substitute Ti than O
    under both Ti-rich and O-rich growth conditions. For S- and P-doped TiO$_2$,
    the doping sites of S and P strongly depend on the preparing method and
    growth condition of the doped
    TiO$_2$.\cite{Yang_2007_JPCC_S,Matsushima_2007_JPCS, Zheng_2010_JPCC} Under
    O-rich growth condition, S (P) prefers to replace Ti and form substitutional
    S (P)-cation doped structure. On the contrary, under Ti-rich growth
    condition, S (P) prefers to replace O and form S (P)-anion doped structure.
    This is consistent with the experiments reported by several independent
    groups.\cite{Yu_2005_EST, Ohno_2004_ACA, Ohno_2003_CL, Umebayashi_2002_APL}
    Yu $\emph{et al.}$ and Ohno $\emph{et al.}$ used the titanium isopropoxide
    and thiourea as the titanium and sulfur original materials, respectively,
    which corresponds to the O-rich growth condition, and prepared the S-cation
    doped TiO$_2$.\cite{Yu_2005_EST, Ohno_2004_ACA, Ohno_2003_CL} And it is
    further confirmed that replacing Ti by S is energetically more favorable
    than replacing O under the O-rich growth condition.\cite{Yu_2005_EST} In
    contrast, Umebayashi $\emph{et al.}$ used TiS$_2$ as the starting material,
    which corresponds to the Ti-rich growth condition, and prepared S-anion
    doped TiO$_2$.\cite{Umebayashi_2002_APL, Umebayashi_2003_CL} Therefore, the
    first-principles theoretical calculations demonstrated a basic experimental fact that the ionic form and site of S (P) dopant in TiO$_2$ can be controlled by the growth condition and preparation method of doped TiO$_2$ sample.\cite{Yang_2007_JPCC_S,Matsushima_2007_JPCS, Zheng_2010_JPCC} For C-doped TiO$_2$, Di Valentin $\emph{et al.}$'s theoretical calculations also gave a similar conclusion.\cite{DiValentin_2005_CM} In addition, the first-principles theoretical calculations for halogen-doped TiO$_2$ further revealed the relationship between the doping sites of halogen atoms and their electronegativities as well as the growth condition of doped TiO$_2$.\cite{Yang_2008_CM, Long_2009_CMS} The following conclusions were drawn\cite{Yang_2008_CM}:

(1) Substitutional X-anion doped TiO$_2$ (X=F, Cl, Br, and I) is energetically preferred to form under the Ti-rich rather than under the O-rich growth condition, and the formation energy increases in the order F $<$ Cl $<$ Br $<$ I. This indicates that it is more difficult to replace an O atom using a larger and less electronegative X atom.

(2) Substitutional X-cation doped TiO$_2$ (X=F, Cl, Br, and I) is energetically preferred to form under the O-rich rather than under the Ti-rich growth condition, and the formation energy increases in the order I $<$ Br $<$ Cl $<$ F. This indicates that it is more difficult to replace a Ti atom using a smaller and more electronegative X atom.

(3) Under O-rich growth condition, it is energetically more favorable to substitute Ti than O using Br and I, while it is energetically more favorable to substitute O than Ti using F and Cl.

(4) Under Ti-rich growth condition, it is energetically more favorable to substitute O than Ti using all the X atoms (X= F, Cl, Br, and I).

\begin{figure}
\center
\includegraphics[scale=0.44]{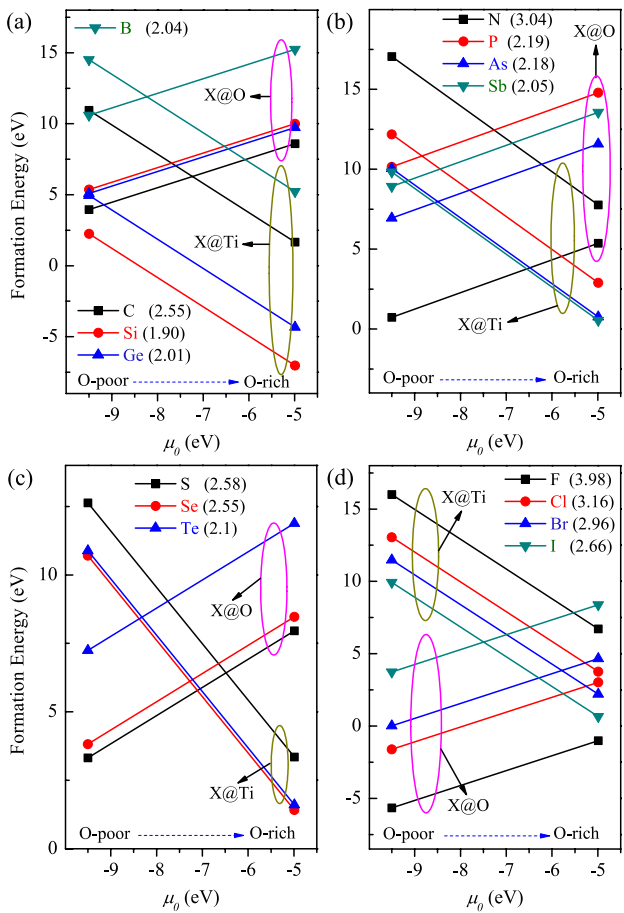} 
\caption{(Color online) Calculated formation energies of substitutional nonmetal-doped TiO$_2$ as a function of the oxygen chemical potential ($\mu _{O}$). (a) B and carbon group (C, Si, and Ge) doped TiO$_2$,
(b) Nitrogen group (N, P, As, and Sb) doped TiO$_2$,
(c) Chalcogen (S, Se, and Te) doped TiO$_2$ and
(d) Halogen (F, Cl, Br, and I) doped TiO$_2$.
X@O and X@Ti represent substitutional X for O and substitutional X for Ti doped structure, respectively.
The Pauling electronegativities of these nonmetal dopants are given inside the parentheses.
}\label{f1}
\end{figure}

To qualitatively show the relationship between the formation energy and the growth condition of doped TiO$_2$ as well as the electronegativities of dopants, we plot the function of the formation energies of nonmetal-doped TiO$_2$ as the oxygen chemical potential (corresponding to the growth condition of TiO$_2$), see Figure \ref{f1}. The following conclusions can be obtained:

(1) For high electronegative F (Cl and N), it is energetically more favorable to substitute O than Ti under both O-rich and Ti-rich conditions.

(2) For low electronegative Si (Ge), it is energetically more favorable to substitute Ti than O under both O-rich and Ti-rich conditions.

(3) For other nonmetal main group elements X- (X=B, C, S, Se, Te, P, As, Sb, Br, and I) doped TiO$_2$, substitutional X-anion doped TiO$_2$ is energetically preferred to form under Ti-rich condition while substitutional X-cation doped TiO$_2$ is energetically preferred to form under O-rich condition. This discrepancy can be partially understood from the electronegativity difference between nonmetal dopants and O (Ti) as well as the bond strength of X-O and X-Ti bonds. For high electronegative F (Cl and N), it is preferred to form F-Ti (Cl-Ti and N-Ti) bond by picking up electrons from Ti rather than form F-O (Cl-O and N-O) bond by losing electrons. For low electronegative Si (Ge), it is preferred to form Si-O (Ge-O) bond by losing electrons to O rather than form Si-Ti (Ge-Ti) bond by picking up electrons. For other nonmetal main group elements X (X=B, C, S, Se, Te, P, As, Sb, Br, and I), their electronegativities are between Ge and N, and thus both X-O and X-Ti bonds are possible to form, depending on the growth conditions of doped TiO$_2$. These results are expected to provide some useful guidance to prepare nonmetal-doped TiO$_2$ and other semiconductor oxides.

Next, let us discuss the possibility of interstitial nonmetal-doped TiO$_2$. Generally speaking, only the dopant with a small atomic size can easily form interstitially doped structure. Therefore, B, C, and N are most possible to be located at interstitial site. First-principles theoretical calculations show that substitutional N-anion and interstitial N-doped structures have nearly the same formation energy under O-rich growth condition.\cite{Yang_2006_JPCB} Di Valentin $\emph{et al.}$ found that interstitial C-doped TiO$_2$ even has lower formation energy than substitutional C-anion doped TiO$_2$ under O-rich growth condition.\cite{DiValentin_2008_CM} Geng $\emph{et al.}$ reported that interstitial B-doped TiO$_2$ has slightly lower formation energy than substitutional B-anion doped TiO$_2$.\cite{Geng_2006_JPCM} Finazzi $\emph{et al.}$ also suggested that the substitutional-anion doped B dopants can be converted into interstitial B after annealing at high temperature on the basis of the first-principles analysis.\cite{Finazzi_2008_JPCC} In summary, for the nonmetal dopants with small atomic size, both the substitutional (at O site) and interstitial structures are possible to form.

\section{Electronic Structure and Optical Absorption}
In 2001, Asahi $\emph{et al.}$ reported the visible-light photocatalytic activity of nitrogen-doped TiO$_2$.\cite{Asahi_2001_Science, Morikawa_2001_JJAP} Soon after, various nonmetal-doped (including B, C, Si, P, S, and halogen elements) TiO$_2$ has been studied to explore their photocatalytic performance under visible-light. Although most of these nonmetal-doped TiO$_2$ show the visible-light optical absorption and photocatalytic activity in some degree, there are some "tricky" fundamental issues in their experimental optical absorption spectra and visible-light photocatalytic mechanism. For example, what leads to the controversy on the origin of the visible-light absorption in N-doped TiO$_2$? Why can B-doped TiO$_2$ show both redshift and blueshift of the optical absorption edge? What is responsible for the different optical absorption thresholds in C-doped TiO$_2$? Which form (anion or cation) of the Si (P and S) ion in TiO$_2$ is more effective to promote the visible-light absorption? Can F-doping lead to the visible-light absorption of TiO$_2$? Why can Cl- and Br-doped TiO$_2$ show the stronger photocatalytic ability than that of undoped TiO$_2$? What is the difference on the mechanism of the visible-light absorption between the I-anion and I-cation doped TiO$_2$? To answer these questions, first-principle theoretical calculation is an effective approach and it has successfully explained many experimental phenomena, including the issues mentioned above. In this section, we review the recent theoretical progress in understanding the electronic structure and optical absorption of nonmetal-doped TiO$_2$ from first-principles calculations.

\subsection{N Doping}
\subsubsection{Origin of Visible-light Absorption}
Since the discovery that N doping can promote the visible-light photocatalytic activity of TiO$_2$,\cite{Asahi_2001_Science, Morikawa_2001_JJAP} numerous experimental efforts have been made to study the N doping influence on the optical absorption and photocatalytic properties of TiO$_2$.\cite{Livraghi_2006_JACS, Irie_2003_JPCB, Lindgren_2003_JPCB, Diwald_2004_JPCB_1, Sakthivel_2004_JPCB, Torres_2004_JPCB, Matsui_2005_JAP, Nakano_2005_APL, Batzill_2006_PRL, Kitano_2006_JPCB, Wang_2006_JMCA, Borras_2007_JPCC, Cong_2007_JPCC, Sathish_2007_ACB, Livraghi_2009_JSSC, Ihara_2003_ACB} With the progress of the theoretical and experimental research, the controversy on the origin of the redshift of the optical absorption edge in N-doped TiO$_2$ appears.\cite{Yang_2007_JPCC_N, Zhao_2008_JPD, Zaleska_2008_RPOE} Three different opinions have been proposed to explain the redshift of the optical absorption edge in N-doped TiO$_2$:

(1) The mixing of N 2$\emph{p}$ states and valence band (VB) leads to the band-gap narrowing. A representative example is Asahi $\emph{et al.}$'s theoretical calculation and experiment,\cite{Asahi_2001_Science, Morikawa_2001_JJAP} and it is further confirmed by the later experiments.\cite{Nakano_2005_APL, Kitano_2006_JPCB, Okato_2005_PRB}

(2) N doping introduces isolated impurity states in the band gap. This is widely confirmed by the experiments\cite{Irie_2003_JPCB, Lindgren_2003_JPCB, Batzill_2006_PRL, Kitano_2006_JPCB} and the first-principles theoretical calculations.\cite{Yang_2006_JPCB, DiValentin_2004_PRB, DiValentin_2005_JPCB, Lee_2005_APL, Xu_2006_JZUSB}

(3) N doping creates oxygen vacancies in TiO$_2$, which introduces defect states in the band gap, and this leads to the visible-light absorption.\cite{Ihara_2003_ACB, Lin_2005_JPCB, Serpone_2006_JPCB, Kuznetsov_2006_JPCB}

In addition, several independent research groups reported that there exists an optimal nitrogen concentration to achieve the highest photocatalytic activity\cite{Kitano_2006_JPCB, Cong_2007_JPCC} or maximize visible-light absorption.\cite{Okato_2005_PRB} It is also found that a band-gap narrowing can be realized in high-concentration N-doped TiO$_2$.\cite{Kitano_2006_JPCB, Okato_2005_PRB} As a consequence, it is speculated that the N-doped TiO$_2$ may show different electronic structure characteristics under different doping concentration, which may lead to the controversy on the origin of the visible-light absorption. This speculation was confirmed through the first-principles electronic structure calculations.\cite{Yang_2006_JPCB}

\begin{figure}
\center
\includegraphics[scale=0.40]{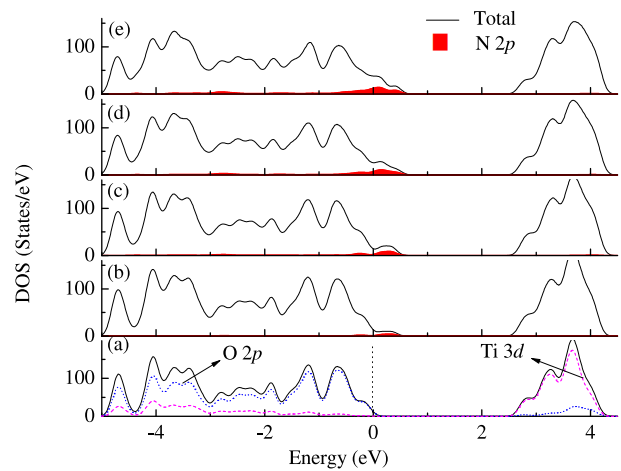}  
\caption{(Color online) Density of states (DOS) of N-anion doped anatase TiO$_2$ with different N-doping levels. (a) Undoped TiO$_2$, (b) 1.04 at.\%, (c) 2.08 at.\%, (d) 3.13 at.\%, and (e) 4.17 at.\%. The energy is measured from the top of the valence band of undoped rutile TiO$_2$.
Reprinted with permission from Yang, K.; Dai, Y.; Huang, B.
\href{http://dx.doi.org/10.1021/jp067491f} {\emph{J. Phys. Chem. C} \textbf{2007}, 111, 12086.} Copyright (2007) American Chemical Society.
}\label{f2}
\end{figure}
\begin{figure}
\center
\includegraphics[scale=0.45]{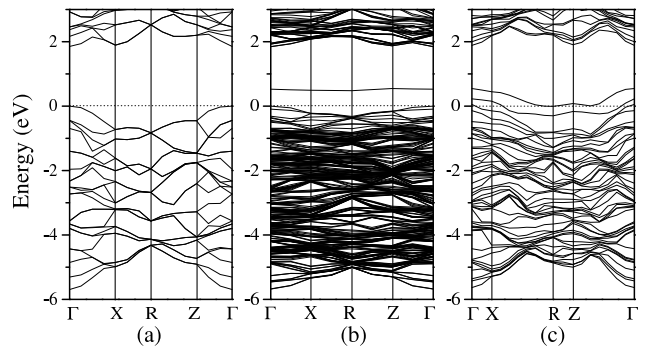} 
\caption{(Color online) Band structure plots for (a) undoped and N-doped rutile TiO$_2$ with a doping level of (b) 1.39 at.\% and (c) 4.17 at.\%. The energy is measured from the top of the valence band of undoped rutile TiO$_2$.
Reprinted with permission from Yang, K.; Dai, Y.; Huang, B. \href{http://dx.doi.org/10.1021/jp067491f} {\emph{J. Phys. Chem. C} \textbf{2007}, 111, 12086.} Copyright (2007) American Chemical Society.
}\label{f3}
\end{figure}

Figure \ref{f2} shows the total density of states (TDOS) and projected density of states (PDOS) of N-doped anatase TiO$_2$ with different N-doping levels; figure \ref{f3} shows the
electronic band structures of undoped and N-doped rutile TiO$_2$ with doping levels at 1.39 at.\% and (c) 4.17 at.\%, respectively. From the electronic structure analysis, it is found that different N-doping levels can lead to different electronic characteristics:\cite{Yang_2007_JPCC_N}

(1) At lower doping concentration ($\leq$ $\sim$2.1 at.\%), N doping introduces some isolated impurity states above the VB in the band gap. They act as transition levels, and thus the electronic transition among the VB, conduction band (CB) and impurity states may be responsible for the redshift of the optical absorption edge.\cite{DiValentin_2004_PRB, DiValentin_2005_JPCB, Lee_2005_APL, Xu_2006_JZUSB}

(2) At higher doping levels ($\geq$ $\sim$4.2 at.\%), more N 2p states are introduced, and they mix with the O 2$\emph{p}$ states, leading to the band-gap narrowing.\cite{Asahi_2001_Science}

Zhao $\emph{et al.}$ also found similar conclusion from first-principles calculations.\cite{Zhao_2008_JPD} Therefore, these results can clarify the controversy on the origin of the redshift of the optical absorption edge in N-doped TiO$_2$, i.e., band-gap narrowing and N 2$\emph{p}$ states in the band gap can be both responsible for the visible-light absorption of N-doped TiO$_2$, which depends on the N doping levels.
 It is noted that same conclusion can be drawn using spin-polarized calculations despite of some differences of the positions of N 2$\emph{p}$ states, which is caused by spin split. It also shows that when nitrogen-concentration exceeds about 2.1 at.\%, the optical energy gap has little further narrowing compared with that at lower doping levels. In contrast, increasing the nitrogen concentration leads to  larger formation energy.\cite{Yang_2007_JPCC_N} This is well consistent with the experiment, in which the higher nitrogen-concentration does not lead to further optical energy-gap narrowing but makes the growth of N-doped TiO$_2$  more difficult.\cite{Okato_2005_PRB} In addition, nonmetal doping (C, S and P) can lower the formation energy of oxygen vacancy,\cite{DiValentin_2005_CM, Graciani_2009_CM, Long_2009_JPCC_S, Long_2009_JPCC_P} and thus the combined effects of the oxygen vacancy and nonmetal doping cannot be excluded to be responsible for the visible-light absorption in N- (C-, S- and P-)doped TiO$_2$.

\subsubsection{Redshift or Blueshift?}

TiO$_2$ has three kinds of crystal phases: anatase, rutile and brookite. Anatase and rutile are two common phases, and thus they are also generally considered to share similar electronic property despite of the different band gap (3.2 eV for anatase phase\cite{Kavan_1996_JACS} and 3.0 eV for rutile phase\cite{Pascual_1977_PRL}). It is known that N doping in TiO$_2$ can induce a redshift of the optical absorption edge.\cite{Asahi_2001_Science, Morikawa_2001_JJAP, Livraghi_2006_JACS, Irie_2003_JPCB, Lindgren_2003_JPCB, Diwald_2004_JPCB_1, Sakthivel_2004_JPCB, Torres_2004_JPCB, Matsui_2005_JAP, Nakano_2005_APL, Batzill_2006_PRL, Kitano_2006_JPCB, Wang_2006_JMCA, Borras_2007_JPCC, Cong_2007_JPCC, Sathish_2007_ACB, Livraghi_2009_JSSC, Ihara_2003_ACB} Surprisingly, Diwald $\emph{et al.}$ observed a blueshift of the optical absorption edge about 0.2 eV in N-implanted rutile TiO$_2$.\cite{Diwald_2004_JPCB_2} Soon after, Di Valentin $\emph{et al.}$ proposed that the band-gap increasing of approximately 0.08 eV in N-doped rutile TiO$_2$ can be responsible for the experimentally observed blueshift on the basis of the first-principles calculations.\cite{DiValentin_2004_PRB} However, this explanation is not in harmony with a basic experimental fact that the redshift is also observed in N-doped rutile TiO$_2$.\cite{Morikawa_2001_JJAP, Diwald_2004_JPCB_1, Torres_2004_JPCB, Batzill_2006_PRL, Livraghi_2009_JSSC} In principle, there are three possible N-doping sites, including substitutional N-anion (N at O site, i.e., N@O), N-cation (N at Ti site, i.e., N@Ti), and interstitial N-doped structures. Hence, to check whether N doping in rutile TiO$_2$ can lead to a blueshift of the optical absorption edge, it is essential to examine the electronic structures of all the possible N-doped TiO$_2$ structures.

\begin{figure}
\center
\includegraphics[scale=0.43]{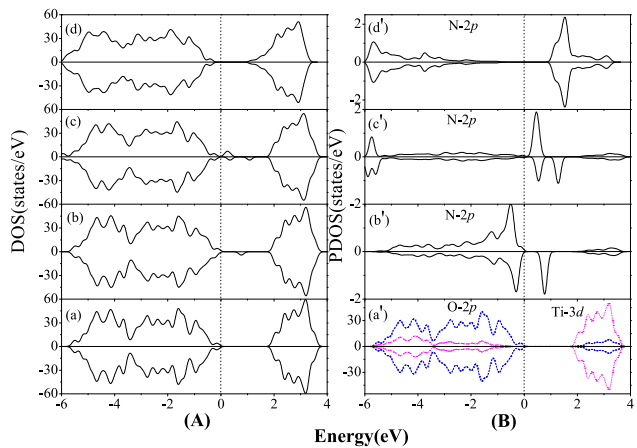} 
\caption{(Color online) TDOS and (B) PDOS of 72-atom rutile supercell. (a) Undoped, (b) substitutional N-anion doped, (c) interstitial N-doped, and (d) substitutional N-cation doped rutile TiO$_2$. The energy is measured from the top of the valence bands of rutile TiO$_2$.
Reprinted with permission from Yang, K.; Dai, Y.; Huang, B.; Han, S. \href{http://dx.doi.org/10.1021/jp0651135} {\emph{J. Phys. Chem. B} \textbf{2006}, 110, 24011.} Copyright (2006) American Chemical Society.
}\label{f4}
\end{figure}

Figure \ref{f4} shows the calculated TDOS and PDOS of three N-doped rutile TiO$_2$ models.\cite{Yang_2006_JPCB}
It shows that N doping either introduces some impurity states in the band gap for substitutional N-anion doped and interstitial N-doped TiO$_2$ or narrows the band gap for substitutional N-cation doped TiO$_2$, which both can lead to a redshift of the optical absorption edge.
 Therefore, the possibility that N doping leads to the blueshift of rutile TiO$_2$ can be excluded. A phase transition from rutile to anatase by nitrogen doping can easily explain the blueshift of the optical absorption edge about 0.2 eV, though Diwald $\emph{et al.}$ did not observe the anatase phase in N-doped rutile TiO$_2$.\cite{Diwald_2004_JPCB_2} In addition, Henderson proposed that the blueshift may be explained according to
 hole trapping effects.\cite{Henderson_2011_SSR}

\subsection{S Doping}
Experiments show that there are two possible substitutional doping sites for S dopants in TiO$_2$, i.e., S anion at O$^{2-}$ site (S@O) and S cation at Ti$^{4+}$ site (S@Ti).\cite{Yu_2005_EST, Ohno_2004_ACA, Ohno_2003_CL, Umebayashi_2002_APL, Umebayashi_2003_CL, Umebayashi_2003_JAP, Ho_2006_JSSC, Li_2007_EST, Sun_2006_IECR, Ohno_2004_WST}, and both the S-anion and S-cation doped TiO$_2$ show the high photocatalytic activity under visible-light. For example, Umebayashi $\emph{et al.}$ prepared S-anion doped anatase and rutile TiO$_2$ by oxidation annealing of titanium disulfide (TiS$_2$) and ion implantation in the rutile single crystal, respectively, and observed the visible-absorption spectrum in these samples.\cite{Umebayashi_2002_APL, Umebayashi_2003_CL, Umebayashi_2003_JAP} Similar visible-light absorption property of S-anion doped TiO$_2$ was also reported by several other groups,\cite{Li_2007_EST, Sun_2006_IECR} and the S doping concentration was found to have a great influence on the visible-light photocatalytic activity of S-anion TiO$_2$.\cite{Ho_2006_JSSC} For substitutional S-cation doped TiO$_2$, interestingly, Yu $\emph{et al.}$ observed a  bactericidal effect under visible light irradiation.\cite{Yu_2005_EST} Ohno $\emph{et al.}$ even found that it exhibits stronger visible-light absorption than N-, C- and the S-anion doped TiO$_2$.\cite{Ohno_2004_ACA, Ohno_2003_CL, Ohno_2004_WST} As in the case of N-doped TiO$_2$, the controversy on the mechanism of the visible-light absorption in S-anion doped TiO$_2$ also exists.
 Earlier theoretical calculations indicated that S-anion doped TiO$_2$ has an obvious band-gap narrowing,\cite{Umebayashi_2002_APL, Umebayashi_2003_JAP, Yamamoto_2004_MT, Sathish_2007_IndianJCA} while the later first-principles calculation shows that S dopants introduce S 3$\emph{p}$ impurity states above the VB, which might be responsible for the redshift of the absorption edge of S-anion doped TiO$_2$.\cite{Tian_2006_JPCB}

\begin{figure}
\center
\includegraphics[scale=0.43]{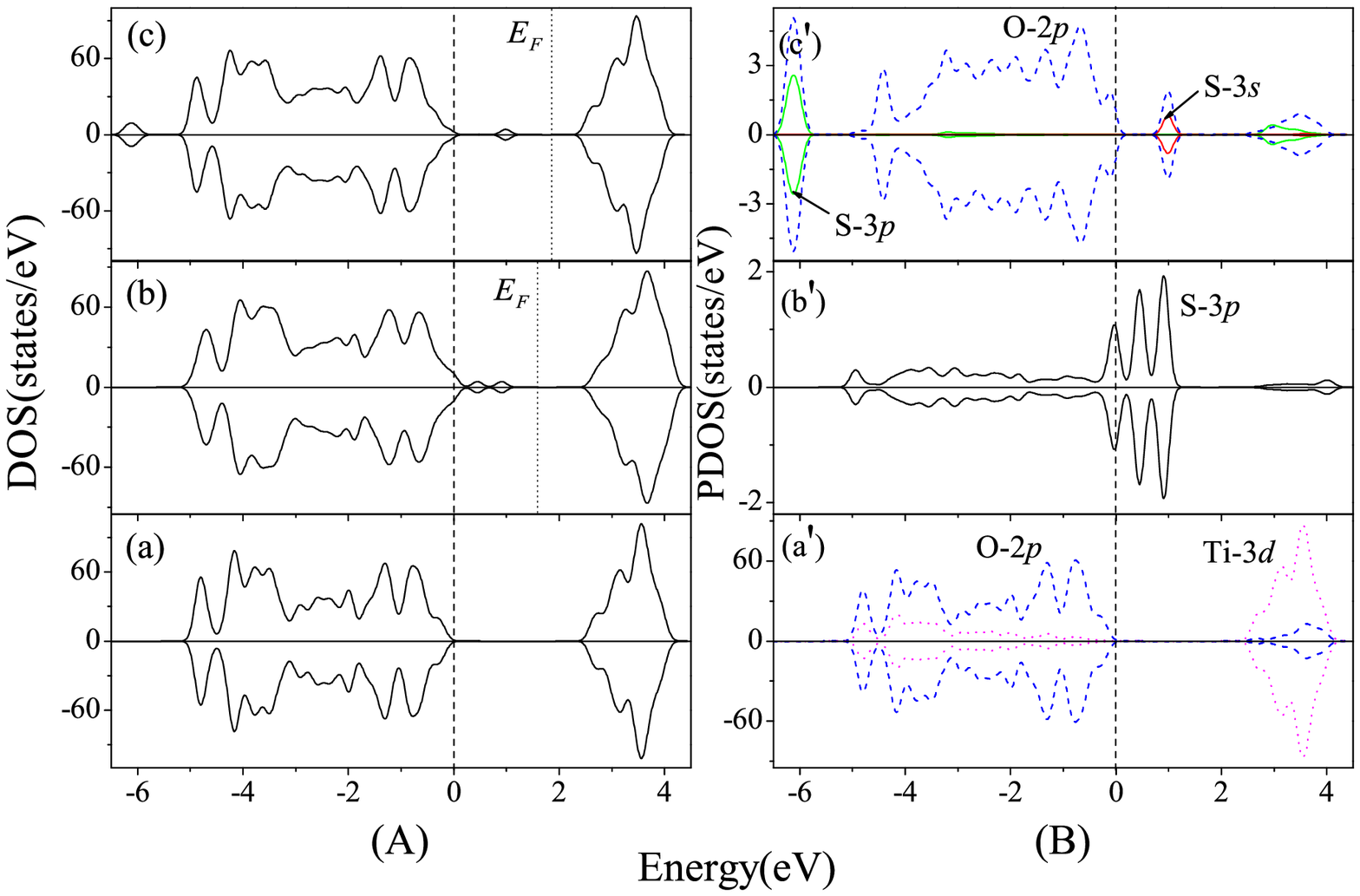} 
\caption{(Color online) (A) TDOS and (B) PDOS of undoped and S-doped anatase TiO$_2$. (a) Undoped, (b) substitutional S-anion (S@O) doped, and (c) substitutional S-cation (S@Ti) doped TiO$_2$. The energy is measured from the top of the valence band of undoped anatase TiO$_2$ and the dot line represents the Fermi level.
Reprinted with permission from Yang, K.; Dai, Y.; Huang, B. \href{http://dx.doi.org/10.1021/jp0756350} {\emph{J. Phys. Chem. C} \textbf{2007}, 111, 18985.} Copyright (2007) American Chemical Society.
}\label{f5}
\end{figure}

\begin{figure}
\center
\includegraphics[scale=0.45]{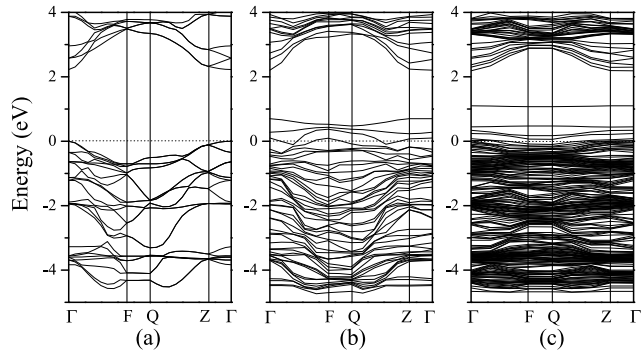}  
\caption{(Color online) Band structure plots for (a) undoped and S-doped anatase TiO$_2$ at (b) 4.17 at.\% and (c) 1.04 at.\% doping level. The energy is measured from the top of the valence band of pure anatase TiO$_2$.
Reprinted with permission from Yang, K.; Dai, Y.; Huang, B. \href{http://dx.doi.org/10.1021/jp0756350} {\emph{J. Phys. Chem. C} \textbf{2007}, 111, 18985.} Copyright (2007) American Chemical Society.
}\label{f6}
\end{figure}

First-principles studies revealed that S-anion and S-cation doped TiO$_2$ show different mechanisms for the visible-light absorption.\cite{Yang_2007_JPCC_S} For S-anion doped TiO$_2$, different doping levels lead to the different electronic structure characteristics. At lower S-doping concentration ($\leq$ 2.08 at.\%), band gap narrows slightly but some S 3$\emph{p}$ localized states are introduced in the band gap (see Figure \ref{f5}b and Figure \ref{f6}c). Therefore, electron excitations from these occupied S 3$\emph{p}$ states to CB might lead to a more significant redshift of the optical absorption edge than the slight band-gap narrowing.\cite{Ho_2006_JSSC} At higher S-doping concentration ($\geq$ 4.17 at.\%), the mixing of the S 3$\emph{p}$ states with the VB causes an obvious band-gap narrowing ($\sim$ 0.7 eV), thus leading to a substantial redshift of absorption spectra (see Figure \ref{f6}b).

For S-cation doped TiO$_2$, S dopants introduce some occupied impurity states consisting of S 3$\emph{s}$ and O 2$\emph{p}$ states in the band gap despite of the unchanged band gap. This indicates that the S dopant has an electron configuration resembling like a S$^{4+}$ (\emph{s$^2$p$^0$})  ion in TiO$_2$ (see Figure \ref{f5}c). Therefore, these impurity states can act as transition levels, and the electron excitations from these transition levels to CB may be responsible for the experimental redshift of the optical absorption edge. Similar S doping influence also appear in rutile TiO$_2$.\cite{Yang_2007_JPCC_S}

\subsection{P Doping}
\begin{figure}
\center
\includegraphics[scale=0.44]{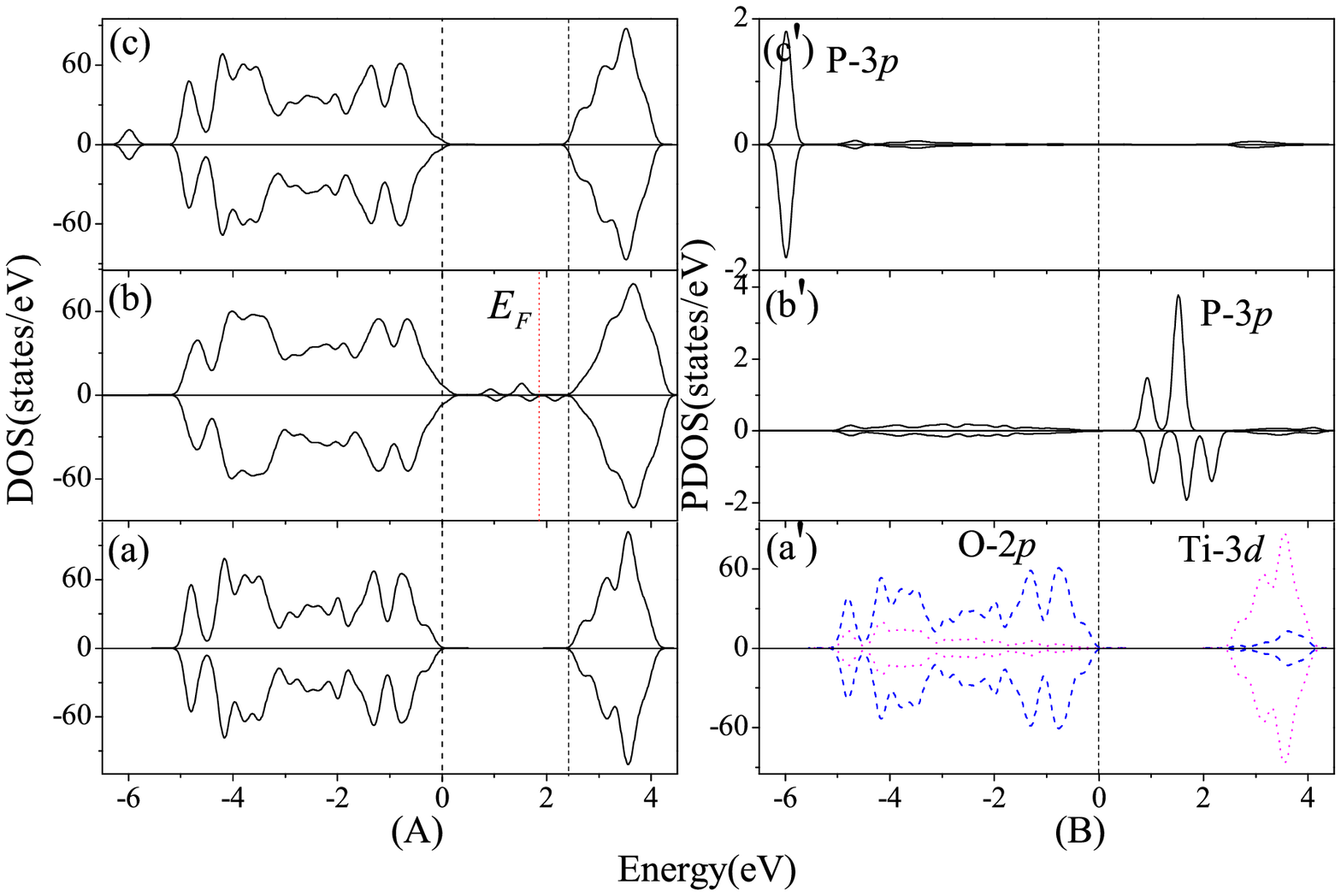} 
\caption{(Color online) (A) TDOS and (B) PDOS of undoped and P-doped anatase TiO$_2$. (a) Undoped, (b) substitutional P-anion (P@O) doped, and (c) substitutional P-cation (P@Ti) doped TiO$_2$. The energy is measured from the top of the valence band of undoped anatase TiO$_2$ and the dot line represents the Fermi level.
Reprinted with permission from Yang, K.; Dai, Y.; Huang, B.
\href{http://dx.doi.org/10.1021/jp0756350} {\emph{J. Phys. Chem. C} \textbf{2007}, 111, 18985.} Copyright (2007) American Chemical Society.
}\label{f7}
\end{figure}

Visible-light photocatalytic activity was also reported in P-anion and P-cation doped TiO$_2$, respectively. Shi $\emph{et al.}$\cite{Shi_2006_JPCB} found that P-cation doped (P@Ti) TiO$_2$ nanoparticles exhibit a stronger visible-light absorption than undoped sample, which is thought to be induced by the impurity states in the band gap. And their X-ray photoelectron spectroscopy (XPS) measurements also indicated that the doped P ions are in the pentavalent-oxidation states (P$^{5+}$).\cite{Shi_2006_JPCB}
Yu $\emph{et al.}$ also found that P-doped TiO$_2$ shows a  better photocatalytic ability than that the pure TiO$_2$.\cite{Yu_2007_JPCS} Li $\emph{et al.}$ observed a visible-light absorption in substitutional P-anion doped (P@O) anatase TiO$_2$, and attributed it to a narrowed band gap.\cite{Lin_2005_CL} On the contrary, Yu $\emph{et al.}$ found a larger band gap in phosphor-modified TiO$_2$ than that of pure TiO$_2$.\cite{Yu_2003_CM} To understand the mechanism underlying these inconsistent experimental observations, first-principles electronic structure calculations for P-anion (P@O) and P-cation (P@Ti) doped TiO$_2$ were performed.\cite{Yang_2007_JPCC_S}
Figure \ref{f7} shows their TDOS and PDOS. P-anion doping does not cause a large band-gap narrowing but introduce P 3$\emph{p}$ states in the band gap. In contrast, P-cation doping neither narrows the band gap nor introduces impurity states in the band gap. The first-principles results indicated
 that the high photocatalytic activity in P-cation doped TiO$_2$ may be caused by the large surface area and the crystallinity of TiO$_2$ instead of the formation of an impurity energy level in the band gap.\cite{Yu_2007_JPCS, Yu_2003_CM}

\subsection{B Doping}
There are three possible doping sites for boron in TiO$_2$, i.e., substitutional B for O (B@O), substitutional B for Ti (B@Ti) and interstitial B site. The atomic radius of B is larger than that of O (0.85 {\AA} $\emph{vs.}$ 0.6 {\AA}) but smaller than that of Ti (0.85 {\AA} $\emph{vs.}$ 1.4 {\AA}). Hence, substitutional B-anion (B@O) doping and interstitial B doping are expected to cause a lattice expansion of TiO$_2$ while substitutional B-cation (B@Ti) doping cause a lattice shrinking. Variable-cell structure optimizations for the three B-doped anatase models were carried out to qualitatively study the B doping influence on the lattice structure of TiO$_2$ .\cite{Yang_2007_PRB} Compared with the undoped anatase model, the volumes of the substitutional B-anion and interstitial B doped models expand about 7.7 \% and 7.0 \%, respectively. In contrast, the volume of substitutional B-cation doped model shrinks about 5 \%. These results are consistent with Jung $\emph{et al.}$'s experiment in which the grain size of anatase phase was enlarged when the incorporation of boron oxides is more than 10\%.\cite{Jung_2004_ACB} However, Chen $\emph{et al.}$ found that the doping of boron ions inhibited the crystal size.\cite{Chen_2006_IECR} Although substitutional B-cation doping can lead to a decrease of the volume of TiO$_2$, this kind of structure is not consistent with Chen $\emph{et al.}$'s experiment in which the B ion is sited at an interstitial position. Therefore, the substitutional B-cation doping cannot account for Chen $\emph{et al.}$'s experimental phenomenon, and further studies are worthy to be done to clarify the boron doping effects on the crystalline size of TiO$_2$.

\begin{figure}
\center
\includegraphics[scale=0.44]{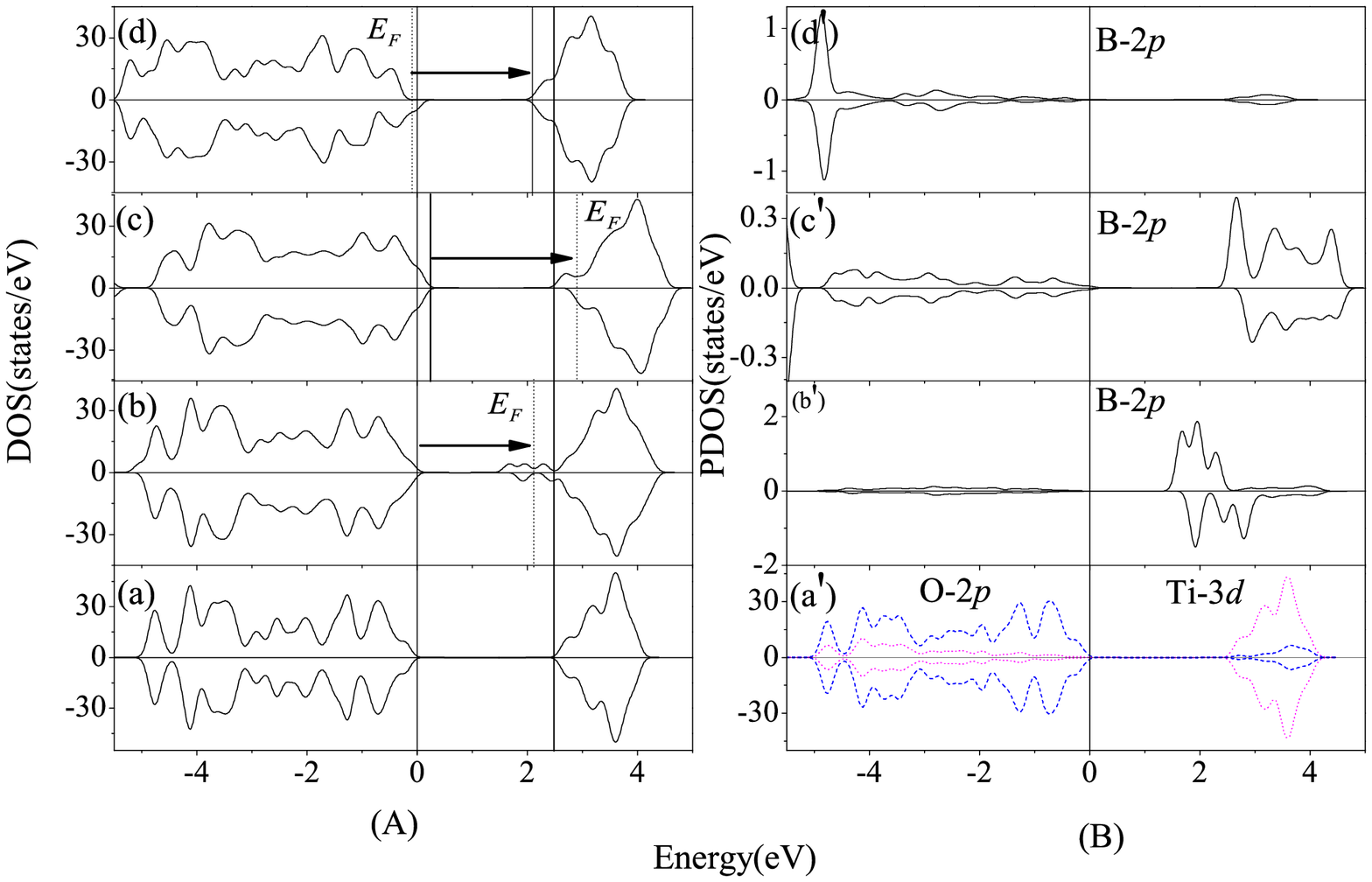} 
\caption{(Color online) (A) TDOS and (B) PDOS of undoped and B-doped TiO$_2$. (a) Undoped, (b) substitutional B-anion (B@O) doped, and (c) substitutional B-cation (B@Ti) doped TiO$_2$. The energy is measured from the top of the valence band of undoped anatase TiO$_2$ and the dot line represents the Fermi level.
Reprinted with permission from Yang, K.; Dai, Y.; Huang, B. \href{http://dx.doi.org/10.1103/PhysRevB.76.195201} {\emph{Phys. Rev. B} \textbf{2007}, 76, 195201.} Copyright (2007) American Physical Society.
}\label{f8}
\end{figure}

Figure \ref{f8} shows the calculated TDOS and PDOS of the three B-doped TiO$_2$ models. For the substitutional B-anion doped TiO$_2$, some impurity states mostly consisting of B 2$\emph{p}$ states are introduced in the band gap.
The electron excitation energy from VB to the unoccupied gap states above the Fermi level decreases about 0.3 eV with respect to the optical band gap of undoped TiO$_2$. This is consistent with the experimental redshift of the optical absorption edge in B-doped TiO$_2$.\cite{Zhao_2004_JACS} It is noted that the substitutional B-anion doping leads to a spin-polarized electron state,\cite{Finazzi_2008_JPCC, Yang_2007_PRB, Zhukov_2010_IJMPB, Yang_2010_JPCC_B} and thus forms a paramagnetic defect consisting of B and adjacent Ti ions.\cite{Finazzi_2008_JPCC} A detailed discussion can be found in section IV. For the substitutional B-cation doped TiO$_2$, its optical band gap decreases about 0.3 eV due to the downward shift of the CB. On the contrary, for the interstitial B-doped TiO$_2$, its optical band gap increases about 0.2 eV for anatase TiO$_2$ and 0.3 eV for rutile TiO$_2$ because of the well-known ``band-filling mechanism"\cite{Pankove_1975_OPS} or ``Moss-Burstein shift"\cite{Moss_1954_PPSB, Burstein_1954_PR}, which is often associated with the optical absorption shift in $\emph{n}$-type semiconductors. This is in good agreement with the experimental blueshift of the optical absorption edge in interstitial B-doped TiO$_2$, in which the optical absorption energy increases about 0.12 eV. Therefore, the blueshift of optical absorption edge in B-doped TiO$_2$ can be attributed to the intrinsic property of interstitial B-doped structure instead of the quantum size effects.\cite{Jung_2004_ACB, Chen_2006_IECR} Similar doping effects also occur in B-doped rutile TiO$_2$.\cite{Yang_2007_PRB, Jin2008JPD} It is also noted that the several different interstitial positions of B dopants should exist,\cite{Finazzi_2008_JPCC} and both the standard and hybrid DFT calculations have been carried out to study the trigonal-planar-coordinated [BO$_3$] and pseudotetrahedral-coordinated [BO$_4$]  species.\cite{Finazzi_2008_JPCC, Feng_2011_JPCC} It is found that the [BO$_3$] specie leads to an increase in the band gap while the [BO$_4$] specie leads to a decrease in the band gap.\cite{Feng_2011_JPCC} However, the [BO$_3$] is shown to be more stable than [BO$_4$].\cite{Finazzi_2008_JPCC}

\subsection{C Doping}
C doping can extend the optical absorption of TiO$_2$ effectively. Interestingly, different degrees of the reduction of the optical band gap have been observed in C-doped TiO$_2$.\cite{Khan_2002_Science, Shen_2006_ML, Hsu_2007_TSF, Wang_2007_ACB, Wong_2007_SCT, Wu_2007_CM, Choi_2004_JMS, Irie_2003_CL, Irie_2006_TSF, Huang_2008_Langmuir, Yang_2007_JC, Mohapatra_2007_JC, Mohapatra_2007_JPCC, Park_2006_NL, Xu_2006_EC} Khan $\emph{et al.}$ firstly found that C-anion-doped rutile TiO$_2$ exhibits two optical absorption thresholds at 535 and 440 nm, which corresponds to the optical band-gap decrease about 0.18 and 0.68 eV, respectively.\cite{Khan_2002_Science} Soon after, more experimental studies confirmed that C-anion doping could induce different degrees of the redshift of the optical absorption edge of TiO$_2$. For example, the optical band-gap reduction of about 0.3 eV and 0.45 eV was observed,\cite{Shen_2006_ML, Hsu_2007_TSF, Wang_2007_ACB, Wong_2007_SCT, Wu_2007_CM} and more pronounced optical band-gap reduction of about 0.72 eV,\cite{Choi_2004_JMS} 0.86 eV,\cite{Irie_2003_CL, Irie_2006_TSF} 0.95 eV,\cite{Huang_2008_Langmuir, Yang_2007_JC} and 1.0 eV \cite{Mohapatra_2007_JC, Mohapatra_2007_JPCC, Park_2006_NL} was also observed. Xu $\emph{et al.}$ found two regions of photo-response from ultraviolet (UV) to 450 nm and 575 nm in C-doped anatase TiO$_2$, which equals to the optical band-gap decrease of 0.45 eV and 1.05 eV, respectively.\cite{Xu_2006_EC} These experiments both indicate that substitutional C-anion doping can lead to several different optical absorption thresholds in TiO$_2$. In addition, carbon can also be incorporated into TiO$_2$ as a cation at a Ti site because of its low electronegativity,
and the C-cation doping influence on the optical absorption spectra of TiO$_2$ is open. Kamisaka $\emph{et al.}$'s first-principle calculations indicate that C-cation doping neither introduces gap states nor induces visible-light absorption.\cite{Kamisaka_2005_JCP} On the contrary, Ren $\emph{et al.}$' observed a visible absorption in the 400-450 nm range in their C-cation doped TiO$_2$ sample using UV-vis diffuse reflectance spectroscopy.\cite{Ren_2007_ACB}

\begin{figure}
\center
\includegraphics[scale=0.36]{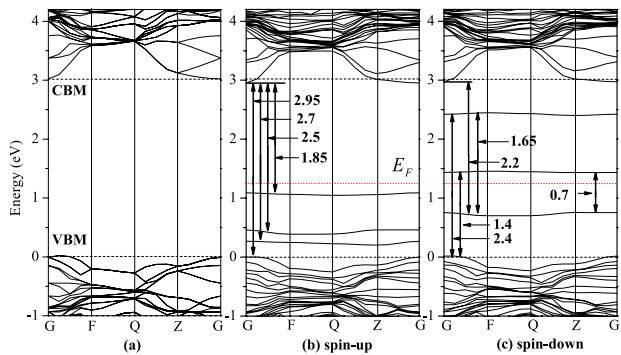} 
\caption{(Color online) Dispersion relations of (a) the non-spin-polarized band of undoped anatase TiO$_2$ and (b, c) the up-spin and down-spin bands of C-cation doped anatase TiO$_2$ with C@O. The red dotted lines represent the Fermi level, and the dashed lines indicate the VBM and the CBM of undoped anatase TiO$_2$.
Reprinted with permission from Yang, K.; Dai, Y.; Huang, B.; Whangbo, M.-H.
\href{http://dx.doi.org/10.1021/jp808483a} {\emph{J. Phys. Chem. C} \textbf{2009}, 113, 2624.} Copyright (2009) American Chemical Society.
}\label{f9}
\end{figure}

Spin-polarized first-principles GGA+\emph{U} calculations were carried to study the electronic and optical absorption properties of C-anion and C-cation doped TiO$_2$.\cite{Yang_2009_JPCC_C} Figure \ref{f9} shows the calculated electronic band structures for undoped and C-anion doped anatase TiO$_2$. With respect to the undoped TiO$_2$, the band gap of C-anion doped TiO$_2$ changes slightly but some spin-polarized impurity states are introduced in the band gap. Correspondingly, associated electron excitations among the VB, the CB and the impurity states can be responsible for the various visible-light absorption thresholds in C anion-doped TiO$_2$. The electron excitation energy from the occupied gap states just above the VB maximum (VBM) to the CB minimum (CBM) decreases about 0.33 - 0.53 eV, which is consistent with the reduction of the optical absorption energy about 0.30 - 0.45 eV.\cite{Shen_2006_ML, Hsu_2007_TSF, Wang_2007_ACB, Wong_2007_SCT, Wu_2007_CM, Xu_2006_EC} The electron excitation energies from the occupied gap states to the CB and from the VB to the empty gap states reduce about 0.63, 0.83 and 1.18 eV. This may be responsible for the large redshift (about 0.70 - 1.05 eV) of the absorption edge in C-doped anatase TiO$_2$ and TiO$_2$ nanotubes.\cite{Choi_2004_JMS, Irie_2003_CL, Irie_2006_TSF, Huang_2008_Langmuir, Yang_2007_JC, Mohapatra_2007_JC, Mohapatra_2007_JPCC, Park_2006_NL, Xu_2006_EC} C-doped rutile TiO$_2$ also shows same electronic characteristics.\cite{Yang_2009_JPCC_C}

\begin{figure}
\center
\includegraphics[scale=0.32]{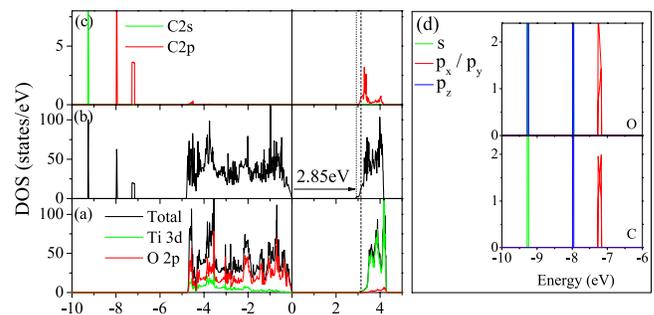} 
\caption{(Color online) (a) Total and projected DOS plots calculated for undoped anatase TiO$_2$. (b) Total DOS plot calculated for C-cation doped anatase TiO$_2$. (c) Projected DOS plots calculated for the C $\emph{s}$ and C 2$\emph{p}$ states of C-cation doped anatase TiO$_2$. (d) Zoomed-in views of the projected DOS plots of the C 2$\emph{s}$/2$\emph{p}$ states and those of the O 2$\emph{s}$/2$\emph{p}$ states for one of the two O atoms forming the linear O-C-O unit. By symmetry, the projected DOS peaks for the 2\emph{p$_x$} and 2\emph{p$_y$} peaks are identical. The energy zero represents the VBM of the undoped and C-cation doped anatase TiO$_2$, the vertical dashed line the CBM of undoped anatase TiO$_2$, and the vertical dotted line the CBM of C-cation doped anatase TiO$_2$.
Reprinted with permission from Yang, K.; Dai, Y.; Huang, B.; Whangbo, M.-H.
\href{http://dx.doi.org/10.1021/jp808483a} {\emph{J. Phys. Chem. C} \textbf{2009}, 113, 2624.} Copyright (2009) American Chemical Society.
}\label{f10}
\end{figure}

Figure \ref{f10} shows DOS plots of undoped and C-cation doped anatase TiO$_2$. For C-cation doped TiO$_2$,it does not show spin-polarized electronic states, and its calculated band gap in the frame of GGA+\emph{U} is about 2.85 eV, less than that of the undoped anatase TiO$_2$ by about 0.18 eV.\cite{Yang_2009_JPCC_C} This can explain the experimental redshift of the optical absorption and enhanced visible-light absorption in the range of 400-450 nm.\cite{Ren_2007_ACB} Similar electronic structure modifications induced by C-cation doping also occur in C-doped rutile TiO$_2$. It is worthy to be mentioned that the local structure around the C dopant in anatase TiO$_2$ significantly influences its electronic properties. Kamisaka $\emph{et al.}$' performed first-principles calculations for C-cation doped TiO$_2$, in which the cell size and shape were fixed, and found that the C dopant forms a planar CO$_3$ species. Neither in-gap impurity states nor visible-light absorbance is found in their calculations.\cite{Kamisaka_2005_JCP} In contrast, Di Valentin $\emph{et al.}$' reported a local structure of CO$_4$ unit and found the band gap of C-cation doped TiO$_2$ shrank slightly about 0.1 eV.\cite{DiValentin_2005_CM} In Yang $\emph{et al.}$'s first-principles calculations, the lattice parameters and all the atomic positions are allowed to relax, in which the doped C atom forms a linear O-C-O unit with a short C-O distance ($\sim$ 1.2 {\AA}) like carbon dioxide O=C=O. Interestingly, this structural characteristic of linear O-C-O unit was confirmed by experimental X-ray Photoelectron Spectroscopy (XPS) study, in which the XPS measurement of the C1$\emph{s}$ binding energy (288.6 eV) shows the presence of C=O bonds.\cite{Ren_2007_ACB}

Interstitial C-doped TiO$_2$ has also been studied using first-principles calculations.\cite{DiValentin_2005_CM, Tian_2010_CPC} It is found that the interstitial C dopants introduce impurity states in the band gap, which can lead to the visible-light absorption in C-doped TiO$_2$.

\subsection{Si Doping}
\begin{figure}
\center
\includegraphics[scale=0.44]{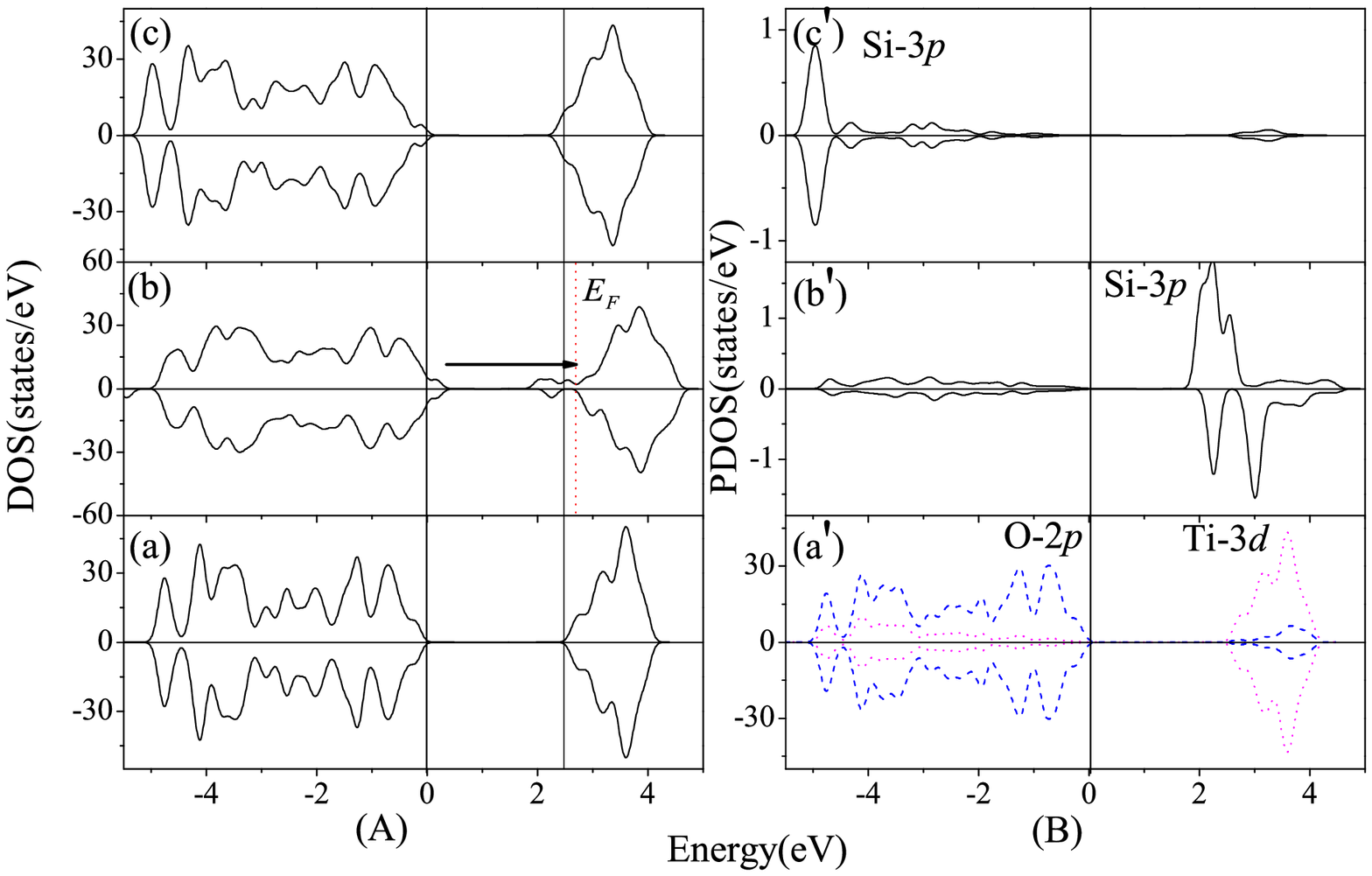}
\caption{(Color online) (A) TDOS and (B) PDOS of undoped and Si-doped anatase TiO$_2$. (a) Undoped, (b) substitutional Si-anion (Si@O) doped, and (c) substitutional Si-cation (Si@Ti) doped TiO$_2$. The energy is measured from the top of the valence band of undoped anatase TiO$_2$ and the dot line represents the Fermi level.
Reprinted with permission from Yang, K.; Dai, Y.; Huang, B. \href{http://dx.doi.org/10.1016/j.cplett.2008.03.018} {\emph{Chem. Phys. Lett.} \textbf{2008}, 456, 71.} Copyright (2008) Elsevier.
}\label{f11}
\end{figure}

Visible-light photocatalytic activity of Si-doped TiO$_2$ has also been reported.\cite{Oh_2003_TSF, Yan_2005_ACB, Ozaki_2005_CL, Ozaki_2007_CataLett} For example, Oh $\emph{et al.}$ found that Si-doping at low level could improve the photocatalytic activity of TiO$_2$, while a high doping over 2\% could decrease its photocatalytic activity .\cite{Oh_2003_TSF} Yan $\emph{et al.}$ observed that substitutional Si-cation doping (Si at Ti site) could cause a redshift of the absorption spectra of TiO$_2$ and favor its photocatalytic activity.\cite{Yan_2005_ACB} Ozaki $\emph{et al.}$ prepared nitrogen-doped silica-modified TiO$_2$, and found a high visible-light photocatalytic activity .\cite{Ozaki_2005_CL, Ozaki_2007_CataLett} Spin-polarized first-principles calculations have been carried out to understand the mechanism of extended visible-light optical absorption.\cite{Yang_2008_CPL} In this work, substitutional Si-anion (Si@O) and Si-cation (Si@Ti) doped TiO$_2$ are modeled, respectively. Figure \ref{f11} shows the calculated
TDOS and PDOS of substitutional Si-doped and undoped anatase TiO$_2$. Compared with the undoped anatase TiO$_2$,substitutional Si-cation doping leads to a band-gap narrowing about 0.25 eV, which is consistent with the experimental visible-light optical absorption.\cite{Oh_2003_TSF, Yan_2005_ACB, Ozaki_2005_CL, Ozaki_2007_CataLett} First-principles calculations also show that high silicon-concentration doping cannot lead to further band-gap narrowing, but requires larger formation energy.\cite{Shi_2011_JSSC}
  For the substitutional Si-anion doped anatase TiO$_2$, visible-light absorption is expected because it shows
    the lower photon absorption energy than that of undoped anatase TiO$_2$.

\subsection{Halogen Doping}
In recent years, many attempts have been made to explore whether halogen doping can improve the visible-light photocatalytic activity of TiO$_2$.\cite{Hattori_2001_JSGST, Wu_2008_Langmuir, Yu_2002_CM, Su_2007_EC, Yamaki_2002_JMSL, Yamaki_2003_NIMB, Li_2005_JFC, Li_2005_CM_1, Li_2005_CM_2, Li_2005_CL, Zhou_2008_JPCC, Huang_2006_JPPA, Huang_2007_JMS, Tang_2007_CM, Luo_2004_CM, Hong_2005_CM, Liu_2006_JPCB} Among all the halogen elements, F is one of the most extensively studied dopants.\cite{Hattori_2001_JSGST, Wu_2008_Langmuir, Yu_2002_CM, Su_2007_EC, Yamaki_2002_JMSL, Yamaki_2003_NIMB, Li_2005_JFC, Li_2005_CM_1, Li_2005_CM_2, Li_2005_CL, Zhou_2008_JPCC, Huang_2006_JPPA, Huang_2007_JMS, Tang_2007_CM} In particular, whether F doping could lead to visible-light absorption is controversial. F-doped TiO$_2$ has shown an improved photocatalytic activity\cite{Hattori_2001_JSGST, Wu_2008_Langmuir} and a stronger absorption in the UV/Vis region, which corresponds to a slight band-gap narrowing about 0.05 eV.\cite{Yu_2002_CM} Similar optical band-gap reduction also occurs in F-doped TiO$_2$ nanotubes\cite{Su_2007_EC} and F-ion-implanted TiO$_2$.\cite{Yamaki_2002_JMSL, Yamaki_2003_NIMB} On the contrary, more experimental studies show that F-doping neither induces any redshifts of the absorption edge of TiO$_2$\cite{Li_2005_JFC, Li_2005_CM_2, Li_2005_CL, Zhou_2008_JPCC} nor changes its optical absorption strength.\cite{Huang_2006_JPPA, Huang_2007_JMS, Tang_2007_CM}
The photocatalytic properties of TiO$_2$ doped with other halogen elements are also reported.Compared with the undoped and F-doped TiO$_2$,Cl and Br co-doped TiO$_2$ exhibits a smaller optical band gap,\cite{Luo_2004_CM} and I-doped TiO$_2$ shows a much better photocatalytic activity than undoped TiO$_2$ under both visible light and UV-visible light.\cite{Hong_2005_CM, Liu_2006_JPCB}

\begin{figure}
\center
\includegraphics[scale=0.49]{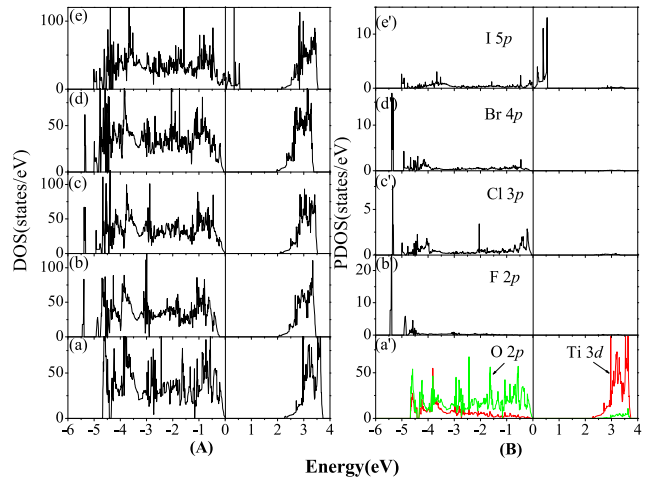} 
\caption{(Color online) TDOS and PDOS plots (left and right panels, respectively) calculated for undoped and X-doped anatase TiO$_2$ with X@O (X = F, Cl, Br, and I): (a) undoped, (b) F-doped, (c) Cl-doped, (d) Br-doped and (e) I-doped anatase TiO$_2$. The energy is measured with respect to the VBM of undoped anatase TiO$_2$.
Reprinted with permission from Yang, K.; Dai, Y.; Huang, B.; Whangbo, M.-H.
\href{http://dx.doi.org/10.1021/cm801741m} {\emph{Chem. Mater.} \textbf{2008}, 20, 6528.} Copyright (2008) American Chemical Society.
}\label{f12}
\end{figure}

To clarify the halogen doping influences on the photocatalytic properties of TiO$_2$ under the UV/Vis light, it is essential to understand whether the doping introduces impurity states in the band gap and how the doping affects the CB and VB edges, i.e., the CBM and VBM of TiO$_2$.
First-principle calculations of halogen-doped TiO$_2$ have been done to understand its electronic properties and origin of the associated visible-light photocatalytic activity.\cite{Yang_2008_CM}
Figure \ref{f12} shows the calculated TDOS and PDOS of the undoped and X-anion doped anatase TiO$_2$. The calculated band gap of the undoped anatase TiO$_2$ is about 2.10 eV. For F-, Cl-, Br- and I-doped TiO$_2$, the calculated band gaps are about 2.06, 1.90, 1.80 and 1.40 eV, respectively. The calculated band gap of the F-doped TiO$_2$ is slightly less than that of the undoped TiO$_2$ (by about 0.04 eV). As a result, it is understandable that either a very slight decrease ($\sim$ 0.05 eV\cite{Yu_2002_CM}) or no optical band gap change was found in experiments.\cite{Li_2005_JFC, Li_2005_CM_2, Li_2005_CL, Zhou_2008_JPCC, Huang_2006_JPPA, Huang_2007_JMS, Tang_2007_CM} It is also noted each F atom ($s^2p^5$) requires one less electron from TiO$_2$ than does each O atom ($s^2p^4$), and thus the substitution of one F$^-$ ion for one O$^{2-}$ introduces one additional electron in the TiO$_2$ lattice. As a result, the treatment of standard DFT calculations for the additional electron leads to the \emph{n}-type conductive property of F-doped TiO$_2$.\cite{Guo_2010_ChinPL} In contrast, in the hybrid DFT or GGA+\emph{U} calculations, the additional electron is shown to reduce one Ti$^{4+}$ (\emph{d$^0$}) to Ti$^{3+}$ (\emph{d$^1$}) ion and introduces localized gap states in the band gap.\cite{Czoska_2008_JPCC, Finazzi_2008_JCP, Long_2010_CPL} However, the lack of the visible-light absorption in F-doped TiO$_2$ indicates that the electron transition from these occupied gap states to the CB is not effective to lead to visible-light absorption.
Therefore, F is not a good dopant to extend the optical absorption edge of TiO$_2$ into the visible light region. In contrast, an obvious band-gap narrowing (about 0.2 and 0.3 eV) is observed in Cl- and Br-doped TiO$_2$, respectively. This is consistent with the experimentally-observed reduction of the band gap by about 0.2-0.3 eV in Cl/Br co-doped TiO$_2$.\cite{Luo_2004_CM} For I-doped TiO$_2$, the band gap narrows about 0.7eV, and thus a significant visible-light optical absorption is expected.

\begin{figure}
\center
\includegraphics[scale=0.44]{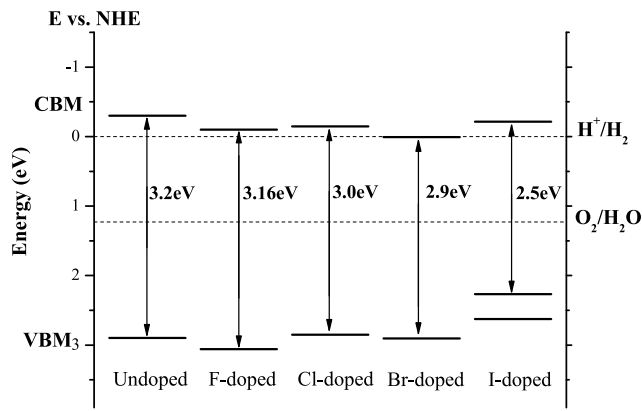} 
\caption{(Color online) Comparison of the calculated VBM and CBM positions of X-doped anatase TiO$_2$ with X@O (X = F, Cl, Br, and I) from the DFT calculations with the corresponding experimental values of undoped TiO$_2$. The VBM and CBM values are given with respect to the normal hydrogen electrode (NHE) potential.
Reprinted with permission from Yang, K.; Dai, Y.; Huang, B.; Whangbo, M.-H.
\href{http://dx.doi.org/10.1021/cm801741m} {\emph{Chem. Mater.} \textbf{2008}, 20, 6528.} Copyright (2008) American Chemical Society.
}\label{f13}
\end{figure}

\begin{figure}
\center
\includegraphics[scale=0.44]{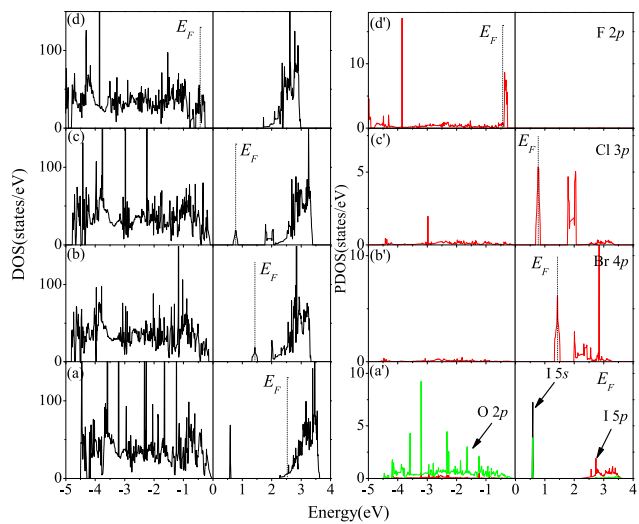} 
\caption{(Color online) TDOS and PDOS plots (left and right panels, respectively) calculated for X-doped anatase TiO$_2$ with X@Ti (X = F, Cl, Br, and I): (a) I-doped, (b) Br-doped, (c) Cl-doped and (d) F-doped anatase TiO$_2$. The energy is measured with respect to the VBM of undoped anatase TiO$_2$.
Reprinted with permission from Yang, K.; Dai, Y.; Huang, B.; Whangbo, M.-H.
\href{http://dx.doi.org/10.1021/cm801741m} {\emph{Chem. Mater.} \textbf{2008}, 20, 6528.} Copyright (2008) American Chemical Society.
}\label{f14}
\end{figure}

The photocatalytic ability of a semiconductor, i.e., the ability to transfer photon-excited electron-hole pairs to the absorbed species on the surface of the semiconductor, can be partially determined by the relative positions of its CBM and VBM with respect to the redox potentials of the adsorbate.\cite{Linsebigler_1995_Chem.Rev} Therefore, the photocatalytic ability of TiO$_2$ can be qualitatively evaluated by the positions of its CBM and VBM. Thermodynamically, the VBM of a photocatalytic semiconductor should lie below the redox level of the adsorbed species while the CBM should lie above the redox level, so that the photoinduced hole in VB can capture an electron from the adsorbed species and the photo-excited electron in CB can be transferred to the adsorbed species. Figure \ref{f13} shows the comparison of the calculated VBM and CBM positions of X-anion doped anatase TiO$_2$ (X=F, Cl, Br, and I) with the experimental values of undoped TiO$_2$. A scissor operation of 1.10 eV was applied to make the calculated band gap comparable to the experimental value. The corrected band gaps for F-, Cl-, Br- and I-doped TiO$_2$ are 3.16, 3.00, 2.90 and 2.50 eV, respectively. For F-doped TiO$_2$, its CBM and VBM shift downwards by 0.20 and 0.16 eV with respect to the undoped TiO$_2$. This indicates that F-doped TiO$_2$ should have a stronger oxidation ability, which can explain the experimentally observed higher photocatalytic activity in F-doped TiO$_2$ than that of the undoped TiO$_2$.\cite{Hattori_2001_JSGST, Wu_2008_Langmuir, Yu_2002_CM, Li_2005_JFC, Huang_2007_JMS, Tang_2007_CM} For Cl-doped TiO$_2$, its CBM shifts downwards by about 0.15 eV while the VBM shifts upwards by 0.05 eV relative to the corresponding values of undoped TiO$_2$. Hence, Cl doping may reduce the ability of oxidation and reduction of TiO$_2$, which is in good agreement with the experimental fact that Cl-doped TiO$_2$ shows lower water-splitting power than that of undoped TiO$_2$ under UV irradiation.\cite{Luo_2004_CM} For Br-doped TiO$_2$, its VBM is nearly same with that of undoped TiO$_2$, and thus Br-doped TiO$_2$ should have the same oxidation ability with that of undoped TiO$_2$. In contrast, the CMB of Br-doped TiO$_2$ shifts downwards by about 0.3eV compared with that of undoped TiO$_2$, so that Br-doped TiO$_2$ might show lower ability to reduce H$^+$ to H$_2$ than does undoped TiO$_2$. For I-doped TiO$_2$, its CBM shifts downward slightly but VBM is raised strongly with respect to that of undoped TiO$_2$, and thus visible-light photocatalytic activity might appear in I-doped TiO$_2$. In summary, the changes of the VBM and CBM can qualitatively explain some experimental facts that the photocatalytic ability was improved upon F, Cl and Br doping.\cite{Hattori_2001_JSGST, Wu_2008_Langmuir, Yu_2002_CM, Li_2005_JFC, Huang_2007_JMS, Tang_2007_CM, Luo_2004_CM}

In addition, the local internal field induced by the dipole moment is considered to be effective to separate the electron-hole pairs and inhibit their recombination,\cite{Sato_2003_JPCB, Long_2006_CPL} and thus the photocatalytic performance of doped TiO$_2$ can be evaluated from the variation of the dipole moment of TiO$_6$ octahedron adjacent to the dopants.\cite{Long_2006_CPL}

Figure \ref{f14} shows the calculated TDOS and PDOS of X-cation (X=F, Cl, Br, and I) doped anatase TiO$_2$. For I-doped anatase TiO$_2$, its VBM and CBM are nearly same with those of undoped TiO$_2$. However, a double-occupied band-gap state above the VBM about 0.6 eV is introduced in the band gap. Its PDOS plot shows that this gap state mostly consists of I 5$\emph{s}$ states and the 2$\emph{p}$ states of the neighboring O atoms around the I dopant, and that I 5$\emph{p}$ states contribute to its CB. This indicates that I dopant exists as an I$^{5+}$ (\emph{s$^2$}) cation, as observed experimentally.\cite{Hong_2005_CM, Liu_2006_JPCB} The Fermi level of I-doped TiO$_2$ is pinned above the CBM about 0.2 eV, indicating that iodine is a good n-type dopant, as in the case of I-doped ZnTe.\cite{Kuhn_1991_SST} Therefore, the optical band gap of I-doped anatase TiO$_2$ decreases about 0.4 eV with respect to that of undoped TiO$_2$. This is mainly responsible for the optical band-gap decrease of I-doped TiO$_2$ and the improved photocatalytic efficiency in the UV/Vis region.\cite{Hong_2005_CM, Liu_2006_JPCB} For substitutional X-cation (X=F, Cl, and Br) doped TiO$_2$, unlike the case of I-doped TiO$_2$, Cl 3$\emph{s}$ and Br 4$\emph{s}$ states do not appear above the VBM. In contrast, singly-filled Cl 3$\emph{p}$ and Br 4$\emph{p}$ states appear in the band gap of Cl- and Br-doped TiO$_2$, respectively. This indicates that Cl and Br at Ti sites exist as Cl$^{4+}$ ($\emph{s$^2$p$^1$}$) and Br$^{4+}$ ($\emph{s$^2$p$^1$}$) ions, respectively. For F-doped TiO$_2$, the F 2$\emph{p}$ states lie in the VB, and the empty F 2$\emph{p}$ states lie just above the Fermi level. Hence, with respect to the Cl- and Br-doped TiO$_2$, the doped F atom at a Ti site should exist as a F$^{3+}$ ($\emph{s$^2$p$^2$}$) ion to a first approximation.\cite{Yang_2008_CM}

\subsection{Hydrogen Impurities in TiO$_2$}
As a ubiquitous impurity, hydrogen (H) widely exists in metal oxides, which forms either deep gap states or shallow donor levels.\cite{Zunger_2002_APL, Peacock_2003_APL, Robertson_2003_TSF, Xiong_2007_JAP} Owning to its small ionic radius, H at an interstitial site is more stable than that at a substitutional site,\cite{Pan_2011_JPCC} though in principle, it can form substitutional structure by replacing the O atom. First-principles calculations show that the interstitial H introduces a shallow donor level in TiO$_2$.\cite{Zunger_2002_APL, Peacock_2003_APL, Robertson_2003_TSF, Xiong_2007_JAP, Park_2006_JKPC, Nahm_2010_JKPC} This is consistent with the recent experiments in which the H is identified as a either a shallow donor or metastable donor in rutile TiO$_2$ single crystal.\cite{Chen_2008_APL, Herklotz_2011_PRB}
Interstitial H introduces an additional electron in TiO$_2$, and thus the \emph{n}-type conductivity can be understandable. However, recent GGA+\emph{U} and hybrid functional calculations show that the additional electrons can be trapped by the Ti ions, reducing Ti$^{4+}$ to Ti$^{3+}$, and introduce the gap states,\cite{Czoska_2008_JPCC, Yang_2010_PRB_TiO} although the position of the gap states strongly depends on the choice of \emph{U} parameters.\cite{Finazzi_2008_JCP}
First-principles calculations also show that substitution of H for O can lead to the \emph{n}-type conductivity, though its structure is not stable.\cite{Pan_2011_JPCC}

It is also interesting to explore the effects of H impurity on the structural stability and electronic property of nonmetal-doped TiO$_2$.\cite{Pan_2011_JPCC, Mi_2007_APL, Li_2011_JAP} With respect to the N-doped TiO$_2$, the combinational doping of (N, H) can enhance the structural stability and lead to a significant band-gap narrowing for anatase and brookite phases of TiO$_2$.\cite{Pan_2011_JPCC, Mi_2007_APL} For N-anion doped TiO$_2$, the doped N dopant exists as N$^{2-}$ ($\emph{s$^2$p$^5$}$) ion, and introduces an unoccupied N 2\emph{p} impurity state in the band gap.\cite{Yang_CPL_2009} For (N, H)-doped TiO$_2$, the interstitial H atom adjacent to the N dopant introduces an additional electron into the lattice, and forms N-H bond by transferring  the additional electron to one empty N 2\emph{p} orbital. As a result, all the N 2\emph{p} impurity states become occupied, leading to either an obvious band-gap narrowing or double-filled N 2\emph{p} impurity states above the VB. This type of charge compensation mechanism also occurs in (N, Ta)- and (N, F)-doped TiO$_2$,\cite{Long_2009_CPL, DiValentin_2008_CM} in which the substitutional Ta-cation (Ta$^{5+}$) and F-anion (F$^-$) doping introduce one additional electron into the TiO$_2$ lattice separately, which fills the empty N 2\emph{p} orbital. For (C, H)-doped TiO$_2$, it is thermodynamically more stable than C-doped TiO$_2$, and a band-gap narrowing also occurs.\cite{Li_2011_JAP} However, in this case, one electron donated by the interstitial H cannot compensate the two holes of C$^{2-}$ ($\emph{s$^2$p$^4$}$) dopant,\cite{Yang_APL_2008} and thus this incomplete charge compensation results in an impurity level just below the CB.\cite{Li_2011_JAP} It is expected that two interstitial H and one substitutional C at an O site can satisfy full charge compensation. In addition, recent experimental and theoretical studies show that hydrogenation on the surface of TiO$_2$ can lead to a structural disorder and introduce mid-gap states, which is responsible for the improved solar-driven photocatalytic activity.\cite{Chen_2011_SCIENCE, Lu_2011_PCCP}
In summary, the influence of hydrogen impurity on the structural stability and electronic property of TiO$_2$ is one interesting topic, and further work is worthy to be done.

\subsection{Co-doping}
Introducing impurity states in the band gap by nonmetal or metal doping is effective to extend the optical absorption edge of TiO$_2$ into the visible-light region, however, the electron-hole recombination center, which generally refers to the impurity-related states near the middle of the band gap, inhibits the photocatalytic efficiency of TiO$_2$. Therefore, narrowing band-gaps without creating mid-gap states is necessary to maximize the photocatalytic performance of TiO$_2$ under the visible-light irradiation.\cite{Yu_2010_JACS, Gai_2009_PRL, Long_2011_PRB} To this aim, co-doping can be an effective approach because it has three important effects:

(1) Co-doping can eliminate impurity states in the band gap via charge compensation between different dopants, which can reduce the number of recombination centers and promote the separation of electron-hole pairs.

(2) Co-doping can facilitate the mixing between the impurity states and VB (CB) by adjusting the position of impurity states in the band gap, and narrow the band gap effectively.

(3) Co-doping can reduce the formation energy of the combination defect with respect to the mono-doped TiO$_2$, and thus improving the solution of ideal dopants in TiO$_2$ or other metal oxides becomes possible.

Herein, we divided the co-doping into three classes, i.e., anion-anion co-doping, anion-cation co-doping and cation-cation co-doping, according to the ionic type (anions or cations) of dopants, and summarized the recent theoretical research progress of the co-doped TiO$_2$, mainly from the viewpoint of charge-compensation.

\subsubsection{Anion-anion Co-doping}
One typical example of anion-anion co-doping is (N, F) co-doped TiO$_2$,\cite{Li_2005_CM_1, DiValentin_2008_CM, Wu_2010_Nanotech, Meng_2009_JMST,Li_2005_CM_2} in which N and F both replace O ions. In N-anion mono-doped TiO$_2$, one N at an O site introduces an accept level in the band gap and exists as a N$^{2-}$ ion, while in F-anion mono-doped TiO$_2$, one F at O site introduces one donor level by donating one additional electron into the TiO$_2$ lattice. As a result, in (N, F) co-doped TiO$_2$, this additional electron can occupy the N 2$\emph{p}$ accept level, and then all the N impurity states become occupied.These occupied N 2$\emph{p}$ states are located just above the VB, and do not act as the electron-hole recombination center, hence improving the visible-light photocatalytic activity of TiO$_2$ with respect to that of the N-anion mono-doped TiO$_2$.\cite{DiValentin_2008_CM, Lu_2012_IJP} In addition, (N, F) co-doping reduces the formation energy with respect to that of the N mono-doped TiO$_2$, and thus a higher nitrogen-concentration doping can be realized. Similar anion-anion charge compensation effect is expected to occur in (N, Cl), (N, Br), (N, I), (P, F), (P, Cl), (P, Br) and (P, I) co-doped TiO$_2$.

\subsubsection{Anion-cation Co-doping}
(N, H$_{int}$) co-doped TiO$_2$ is a typical example of the anion-cation co-doping, in which N replaces O while H locates at an interstitial site.  In this system, one electron donated by an interstitial H occupy the N 2$\emph{p}$ acceptor level, and the band gap narrows about 0.26 eV, much larger than the N-anion mono-doped TiO$_2$ (0.04 eV).\cite{Mi_2007_APL}

Another two kinds of anion-cation co-doping combinations are (N, X) and (C,Y),\cite{Long_2009_CPL,Gai_2009_PRL,Obata_2007_CP,Long_2010_CM} in which X can be Ta$^{5+}$ or Nb$^{5+}$ while Y\footnote{Herein Y is not element yttrium.} can be W$^{6+}$ or Mo$^{6+}$. It is known that N dopant exists as N$^{2-}$ ($\emph{s$^2$p$^5$}$)  ion at an O site,\cite{Yang_CPL_2009} and introduces one acceptor level in the band gap, while pentavalent Ta (Nb) cation at a Ti site donates one electron into TiO$_2$ lattice, introducing one donor level in the band gap. As a result, the charge compensation can occur in (N, Ta),\cite{Long_2009_CPL,Obata_2007_CP} and (N, Nb) co-doped TiO$_2$, and the electron-hole recombination center (i.e., the impurity-related states near the middle of the band gap) might be removed, thus promoting the rate of electron-hole separation and improving its photocatalytic efficiency. Similarly, the C dopant exists as C$^{2-}$ ($\emph{s$^2$p$^4$}$) ion at an O site,\cite{Yang_APL_2008} and introduces two acceptor levels in the band gap. In contrast, the hexavalent W (Mo) cation at a Ti site donates two electrons into TiO$_2$ lattice, i.e., introducing two donor levels. Therefore, in (C, W) and (C, Mo) co-doped TiO$_2$,\cite{Gai_2009_PRL, Long_2010_CM} the two acceptor levels introduced by C dopant become occupied by capturing the two electrons donated by W and Mo, and C 2$\emph{p}$ impurity states are more close to the VB, reducing the number of electron-hole recombination centers. However, it is noted that the full charge compensation only occurs when the doped anion and cation are bonded together, though this configuration generally possesses the lowest total energy.\cite{Long_2010_CM, Liu_2011_ASS} This is reasonable because the doped anion and cation can form a strong bond by a direct charge transfer.

\begin{figure}
\center
\includegraphics[scale=0.23]{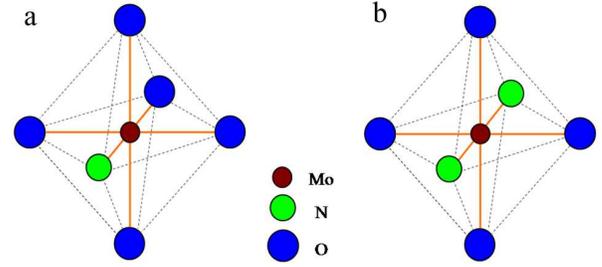}
\caption{(Color online) Relative positions of substitutional N anion and Mo cation in TiO$_2$: (a) Non-passivated doping, and (b) passivated doping.
Reprinted with permission from Liu, H.; Lu, Z.; Yue, l.; Liu, J.; Gan, Z.; Shu, C.; Zhang, T.; Shi, J.; Xiong, R.
\href{http://dx.doi.org/10.1016/j.apsusc.2011.05.085} {\emph{Appl. Sur. Sci.} \textbf{2011}, 257, 9355.} Copyright (2011) Elsevier.
}\label{f15}
\end{figure}

\begin{figure}
\center
\includegraphics[scale=0.3]{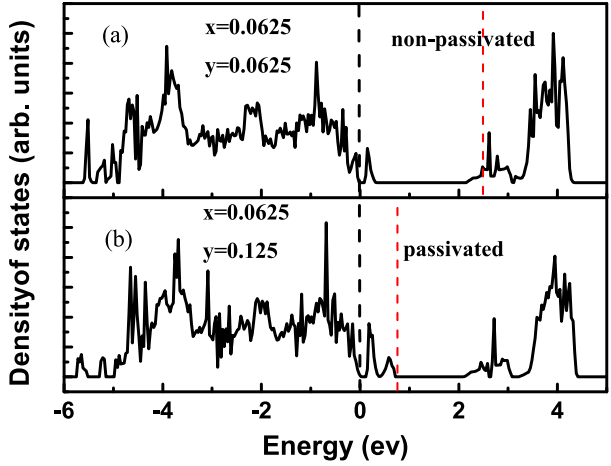}
\caption{(Color online) TDOSs of (Mo+N) co-doped TiO$_2$ (Ti$_{1-x}$Mo$_x$O$_{2-y}$N$_y$): (a) Non-passivated and (b) passivated. The red dash lines represent the Fermi level.
Reprinted with permission from Liu, H.; Lu, Z.; Yue, l.; Liu, J.; Gan, Z.; Shu, C.; Zhang, T.; Shi, J.; Xiong, R.
\href{http://dx.doi.org/10.1016/j.apsusc.2011.05.085}
{\emph{Appl. Sur. Sci.} \textbf{2011}, 257, 9355.} Copyright (2011) Elsevier.
}\label{f16}
\end{figure}

Recently, another kind of anion-cation co-doping such as (N, Mo) co-doping has been proposed,\cite{Liu_2011_ASS} in which two N anions are bonded to one hexavalent Mo cation, forming a near-linear N-Mo-N unit. For convenience, we refer it to as (N-Mo-N) co-doping. In this case, two electrons donated by one hexavalent Mo cation at a Ti site can compensate two holes introduced by two N anions, thus achieving full charge compensation. Liu $\emph{et al.}$ further studied the difference of the electronic structures between non-passivated and passivated (N, Mo) co-doping.\cite{Liu_2011_ASS} The local geometrical configurations of N and Mo dopants for these two kinds of (N, Mo) co-doped TiO$_2$ are shown in Figure \ref{f15}. Their calculated TDOS and PDOS plots are shown in Figure \ref{f16}. For non-passivated (N, Mo) co-doping, one N anion is bonded to one Mo cation, and the doping ratio between N and Mo is 1. In this case, one hole introduced by one N cannot compensate two electrons donated by one Mo, thus leading to \emph{n}-type conductivity. In passivated (N, Mo) co-doped TiO$_2$, i.e., (N-Mo-N) co-doped system, the two electrons are fully compensated by the two holes introduced by the two N anions, and the N 2$\emph{p}$ acceptor levels are removed and do not act as an electron-hole recombination centers, thus improving the photocatalytic efficiency.

\begin{figure}
\center
\includegraphics[scale=0.44]{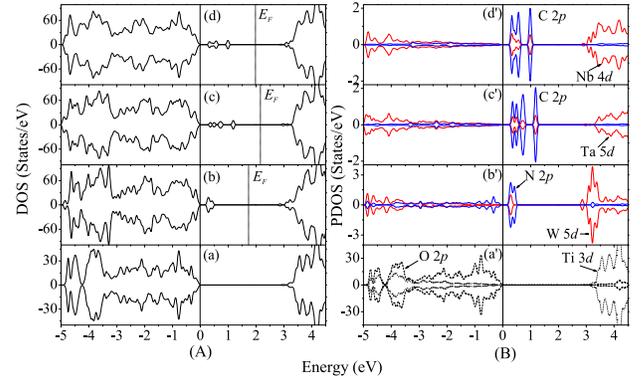}
\caption{(Color online) TDOS and PDOS plots (left and right panels, respectively) calculated for undoped and anion-cation co-doped anatase TiO$_2$: (a) Undoped, (b) (N-W-N) co-doped, (c) (Ta-C-Ta) co-doped and (d) (Nb-C-Nb) co-doped anatase TiO$_2$. The VBM of each compound was set to zero for a convenient comparison.
}\label{f17}
\end{figure}

Similar to the case of (N-Mo-N) co-doped TiO$_2$, we speculate that the full charge compensation can also occur in (N-W-N), (Ta-C-Ta), and (Nb-C-Nb) co-doped TiO$_2$. The calculated TDOS and PDOS using GGA+\emph{U} method for these three co-doped TiO$_2$ are shown in Figure \ref{f17}. The details of \emph{U} value setting can be found in previous work.\cite{Yang_2010_PRB_TiO} For comparison, the TDOS and PDOS for the undoped TiO$_2$ are also shown. For (N-W-N) co-doped TiO$_2$, the two electrons donated by the W cation compensate the two holes introduced by the two N anions. With respect to the N-anion mono-doped TiO$_2$, all the N 2$\emph{p}$ states are occupied and lie just above the VB in this system, and no accept levels exist in the middle of the band gap. For (Ta-C-Ta) and (Nb-C-Nb) co-doped TiO$_2$, the two electrons donated by two Ta (Nb) electrons compensate the two holes introduced by one C dopant, and all the C 2$\emph{p}$ states become occupied. Compared with that of (Ta-C-Ta) co-doped TiO$_2$, the C 2$\emph{p}$ states in (Nb-C-Nb) co-doped TiO$_2$ is much close to the VB, indicating that the C dopant forms a stronger C-Nb bond than C-Ta bond. It is also noted that, with respect to C- or N-anion mono-doped TiO$_2$,\cite{Yang_CPL_2009,Yang_APL_2008} the three co-doped systems become non-spin-polarized because of the full charge compensation.

\subsubsection{Cation-cation Co-doping}
\begin{figure}
\center
\includegraphics[width=3.2in,height=2.1in]{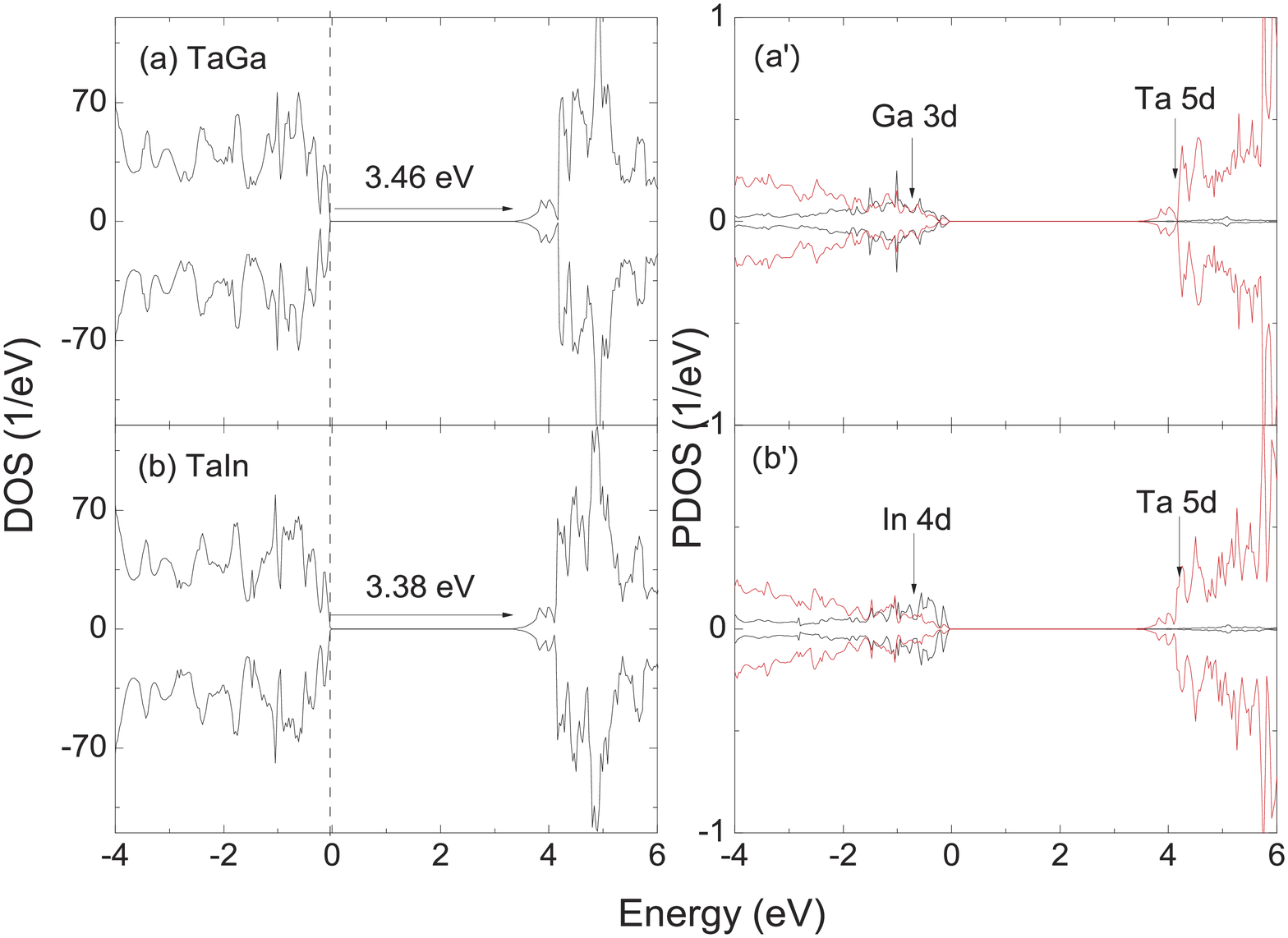}
\caption{(Color online) Total DOS and PDOS plots for (Ta, Ga/In) co-doped 108-atom supercell anatase. The top of the valence band of pure anatase is set as zero.
Reprinted with permission from Long, R.; English, N. J. \href{http://dx.doi.org/10.1103/PhysRevB.83.155209} {\emph{Phys. Rev. B} \textbf{2011}, 83, 155209.} Copyright (2011) American Physical Society.
}\label{f18}
\end{figure}

\begin{figure}
\center
\includegraphics[width=3.2in,height=2.1in]{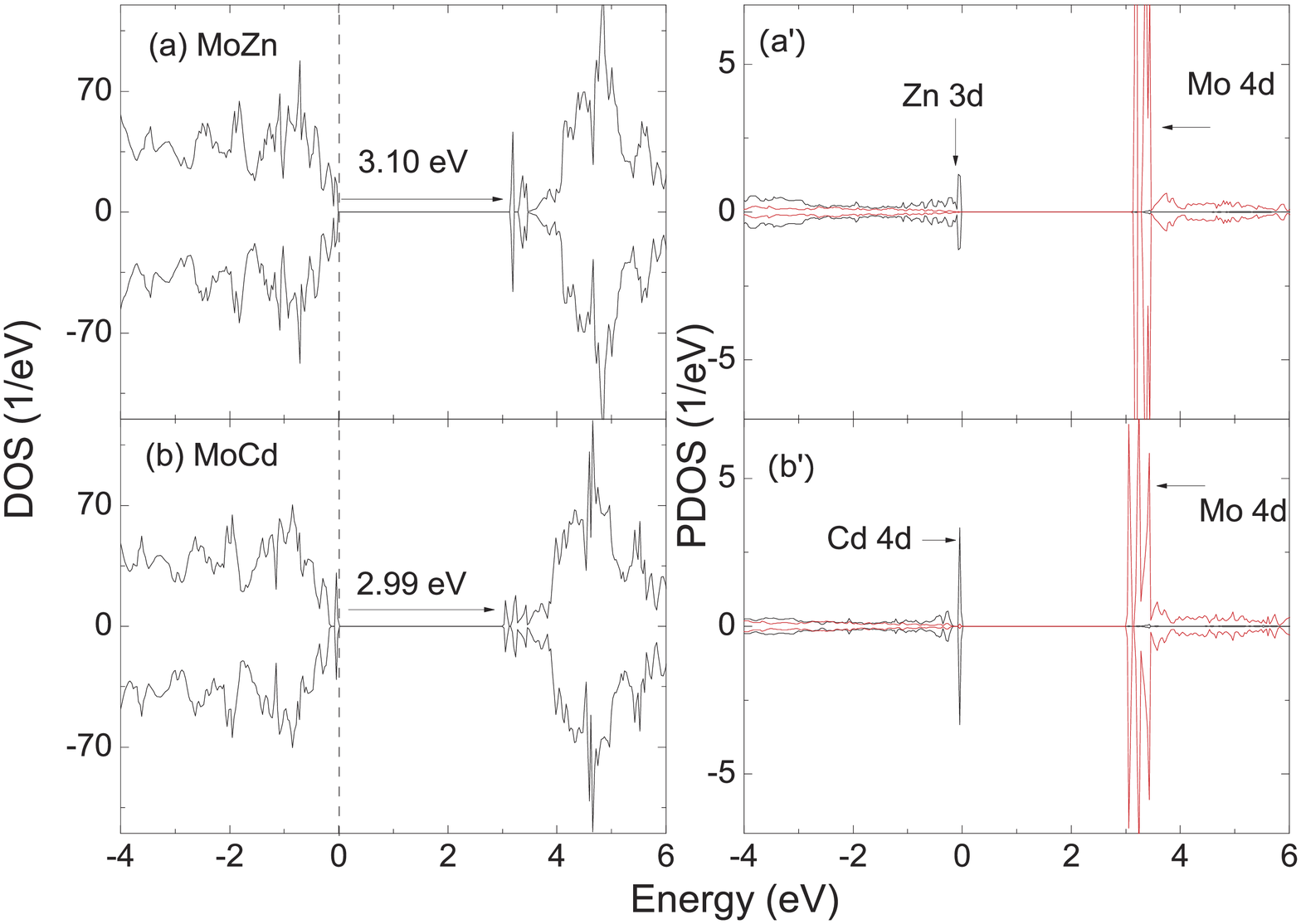}
\caption{(Color online) Total DOS and PDOS plots for (Mo, Zn/Cd) co-doped 108-atom supercell anatase. The top of the valence band of pure anatase is set as zero.
Reprinted with permission from Long, R.; English, N. J. \href{http://dx.doi.org/10.1103/PhysRevB.83.155209} {\emph{Phys. Rev. B} \textbf{2011}, 83, 155209.} Copyright (2011) American Physical Society.
}\label{f19}
\end{figure}

Cation-cation co-doping has also been proposed to narrow the band gap of TiO$_2$, and two criteria of choosing cations are suggested.\cite{Long_2011_PRB} Firstly, the cation should have a closed-shell electronic configuration like $\emph{d$^0$}$ or $\emph{d$^{10}$}$. Secondly, to keep the semiconductor characteristic of TiO$_2$, co-doping with cations A$^{x+}$ and B$^{y+}$ should satisfy a simple rule, i.e., $\emph{x}$ + $\emph{y}$ = 8. On the basis of the criteria, the cation-cation co-doping combination of (Mo$^{6+}$, Zn$^{2+}$/Cd$^{2+}$) and (Ta$^{5+}$, Ga$^{3+}$/In$^{3+}$) is expected to cause an effective band-gap narrowing without introducing gap states. In these two kinds of cation-cation co-doped TiO$_2$, the number of the holes introduced by Zn$^{2+}$/Cd$^{2+}$ (Ga$^{3+}$/In$^{3+}$) equals to that of electrons introduced by Mo$^{6+}$ (Ta$^{5+}$), and thus full charge compensation can be obtained. In addition, the electronic configuration of $\emph{d$^0$}$ or $\emph{d$^{10}$}$ of the cations can guarantee that the $\emph{d}$ states of dopants either are fully unoccupied or occupied, and thus no \emph{d}-orbital-related impurity states can be created in the band gap.
Figure \ref{f18} and Figure \ref{f19} show the calculated TDOS and PDOS of (Ta$^{5+}$, Ga$^{3+}$/In$^{3+}$) and (Mo$^{6+}$, Zn$^{2+}$/Cd$^{2+}$) co-doped TiO$_2$, respectively. As shown from the plots, the charge compensation occurs in the two systems, and with respect to cation mono-doped TiO$_2$, the mid-gap states are passivated.

In summary, co-doping can be an effective approach to narrow the band gap of semiconductor photocatalyst and get rid of the electron-hole recombination centers through the full charge compensation. To this aim, the number of the holes on acceptor levels should be equal to that of the electrons on donor levels. Therefore, we can define the following equation to be a basic rule for co-doping:
\begin{equation}
N_a \times N_h = N_d \times N_e.
\end{equation}
$N_a$ and $N_d$ represent the number of acceptor atoms and donor atoms, respectively. $N_h$ represents the number of introduced holes per acceptor atom and $N_e$ represents the number of donated electrons per donor atom. It is noted that the Ti (O) ions in TiO$_2$ and SrTiO$_3$ have same chemical state, and thus the co-doping combinations mentioned above can also be applied in SrTiO$_3$.\cite{Wei2010PCCP, Sulaeman2010_IJOPT}
For a reference to choose an ideal combination of dopants for co-doping in TiO$_2$ and SrTiO$_3$, we list the common acceptor (donor) atoms as well as the number of the holes (electrons) per acceptor (donor) atom in Table \ref{t1}.

\begin{table}
\centering
\caption{
List of the acceptor atoms and donor atoms in TiO$_2$. The numbers in the third column represents the introduced holes per acceptor atom ($N_h$) or donated electrons per donor atom ($N_e$). H$_{int}$ and Li$_{int}$ represent the interstitial H and Li ions, respectively.}
{\footnotesize
\begin{tabular} {c|c|c|c}
\hline
\multicolumn{2}{c} {\textbf{Acceptor atom}} & $N_h$ &  \\
\hline
Anion	&	N$^{2-}$, P$^{2-}$	&	1	&	 \cite{Yang_CPL_2009,Yang_2007_JPCC_S}	 \\
	&	C$^{2-}$	&	2	&	\cite{Yang_APL_2008}	\\
	&	B$^{2-}$	&	3	&	\cite{Yang_2010_JPCC_B}	\\
\hline
Cation	&	B$^{3+}$, Al$^{3+}$, Ga$^{3+}$, In$^{3+}$ 	&	1	 &	 \cite{Yang_2010_JPCC_B}	\\
	&	Sc$^{3+}$, Y$^{3+}$, La$^{3+}$	&	1	& \textsuperscript{\emph{a}}\\
	&	Be$^{2+}$, Mg$^{2+}$, Ca$^{2+}$, Sr$^{2+}$, Ba$^{2+}$	 &	2	&	 \cite{Yang_2010_JPCC_B}	\\
	&	Li$^+$, Na$^+$, K$^+$, Rb$^+$, Cs$^+$	&	3	&	 \cite{Yang_2010_JPCC_B,Tao_2010_PLA}	\\
\hline
\multicolumn{2}{c} {\textbf{Donor atom}}  & $N_e$ & \\
\hline
Anion	&	F$^-$, Cl$^-$, Br$^-$, I$^-$	&	1	&	 \cite{Yang_2008_CM,Czoska_2008_JPCC}	\\
\hline
Cation	&	H$_{int}$, Li$_{int}$	&	1	&	 \cite{Mi_2007_APL,Zunger_2002_APL,Peacock_2003_APL,Robertson_2003_TSF}	 \\
	&	I$^{5+}$, Nb$^{5+}$, Ta$^{5+}$	&	1	&	 \cite{Yang_2008_CM,Long_2009_CPL}	\\
	&	Mo$^{6+}$, W$^{6+}$	&	2	&	\cite{Long_2011_PRB}	 \\
\hline
\end{tabular}
}
\label{t1}

\textsuperscript{\emph{{\footnotesize a}}} {\footnotesize Our DFT calculations indicated that the substitutional Sc (Y and La)-cation doping introduces one hole in O 2\emph{p} orbital, and produces a spin magnetic moment of 1.0 {$\mu$}$_B$.}
\end{table}

\section{Magnetic Properties of C-, N- and B-doped TiO$_2$}
As one of the widely used wide-band-gap semiconductors, TiO$_2$ has also attracted great interest owning to its potential applications in the spintronics. Since the discovery of the room-temperature ferromagnetism (RTFM) in Co-doped TiO$_2$,\cite{Matsumoto_2001_SCIENCE} lots of experimental attempts have been made to obtain a stable RTFM through doping TiO$_2$ and ZnO with various transition-metals (TMs).\cite{Hong_2004_APL, Shut_2006_nimb, Kim_2004_APL, Duhalde_2005_PRB, Kitt_2006_PRL, Sluiter_2005_PRL} Recently, unexpected RTFM has been observed in two classes of semiconductors in the absence of TM dopants. One class is undoped oxides such as CaO,\cite{Elfimov_2002_PRL} HfO$_2$,\cite{Venkatesan_2004_Nature} TiO$_2$,\cite{Hong_2006_PRB} ZnO\cite{Hong_2007_JPCM} and SnO$_2$,\cite{Hong_2008_PRB} the ferromagnetism of which is attributed to the cation vacancy.\cite{Elfimov_2002_PRL, Pemmaraju_2005_PRL, Peng_2009_PRL, Peng_2009_PRB, Yang_2010_PRB_TiO} The other one is doped oxides with 2$\emph{p}$ light elements, such as C (N)-doped ZnO,\cite{Pan_PRL_2007, Yu_2007_JPD, Zhou_2008_APL} TiO$_2$,\cite{Wen_IEEE.Trans.Magn_2009, Ye_PLA_2009, Cruz_JPCM_2009}  SrO,\cite{Elfimov_2007_PRL} SnO$_2$,\cite{Hong_APL_2011} In$_2$O$_3$,\cite{Ruan_2010_SSC, Khan_2011_JMMM} SrTiO$_3$,\cite{Liu_CJP_2009} and BaTiO$_3$.\cite{Tan_JAC_2011} The ferromagnetism in the two types of materials is also called as ``$\emph{d$^0$}$" magnetism,\cite{Coey_2005_SSS} because the magnetism is not caused by the partially filled $\emph{d}$ electrons. Soon after the experimental discovery of the RTFM in C-doped ZnO,\cite{Pan_PRL_2007} lots of theoretical efforts have been made to explore the origin of the $\emph{d$^0$}$ magnetism in 2$\emph{p}$-light-element doped oxides, sulfides and nitrides.\cite{Yang_APL_2008, Yang_CPL_2009, Yang_2010_JPCC_B, Yang_PRB_2010, Shen_PRB_2008, Peng_2009_APL, Long_2009_PRB, Fan_2009_APL, Wang_2009_PLA, Chen_2011_PLA, Yang_2011_APL} Here we review the theoretical progress of the $\emph{d$^0$}$ magnetism in C-, N- and B- doped TiO$_2$.

\subsection{Origin of Spin-polarization}

\begin{figure}
\center
\includegraphics[scale=0.45]{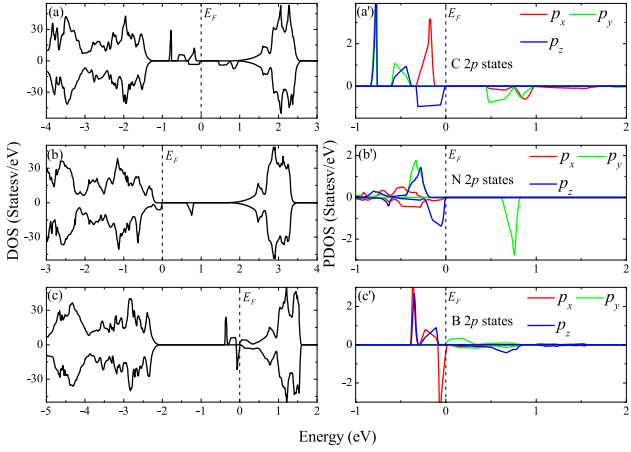}  
\caption{(Color online) Total DOS and PDOS plots for (a) C-doped, (b) N-doped and (c) B-doped anatase TiO$_2$. The Fermi level is indicated by the dashed line at 0 eV.
}\label{f20}
\end{figure}

\begin{figure}
\center
\includegraphics[scale=0.45]{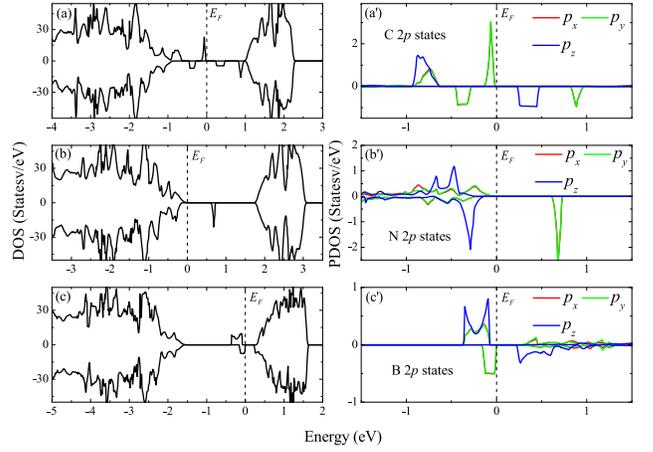} 
\caption{(Color online) Total DOS and PDOS plots for (a) C-doped, (b) N-doped and (c) B-doped rutile TiO$_2$. The
Fermi level is indicated by the dashed line at 0 eV.
}\label{f21}
\end{figure}

The spin-polarization in substitutional C-, N- and B-doped TiO$_2$ has been studied.\cite{Yang_APL_2008, Yang_CPL_2009, Yang_2010_JPCC_B, Wang_2009_SSC, Li_2007_PSSPRL, Tao_APL_2009, Bai_2008_SSC} It is useful to analyze the mechanism of the spin-polarization by examining the PDOS of the dopant. Figure \ref{f20} shows the TDOS and PDOS of C- (N- and B-) doped anatase TiO$_2$. It clearly shows that the substitution of C (N and B) for O leads to a substantial spin-polarization. For C-doped TiO$_2$, the up-spin C 2$\emph{p$_x$}$, 2$\emph{p$_y$}$ and 2$\emph{p$_z$}$ orbitals are both occupied, and the down-spin C 2$\emph{p$_z$}$ orbital is also occupied, but the down-spin C 2$\emph{p$_x$}$ and 2$\emph{p$_y$}$ orbitals are unoccupied. This indicates that one doped C atom at an O site generates a magnetic moment of 2.0 {$\mu$}$_B$, and has an electron configuration like that of a C$^{2-}$ ($\emph{s$^2$p$^4$}$) anion. For N-doped TiO$_2$, N has one more electron than C, and the additional electron occupied the down-spin N 2$\emph{p$_x$}$ orbital, thus leading to a magnetic moment of 1.0 {$\mu$}$_B$. The doped N shows the electron configuration of N$^{2-}$ ($\emph{s$^2$p$^5$}$) anion. For B-doped TiO$_2$, B has one fewer electron than C, and thus it should exist as a B$^{2-}$ ($\emph{s$^2$p$^3$}$) anion at an O site. Its PDOS shows that up-spin and down-spin 2$\emph{p$_x$}$ orbitals and up-spin 2$\emph{p$_z$}$ orbital are occupied, but the other three orbitals are unoccupied, thus resulting in a magnetic moment of 1.0 {$\mu$}$_B$. Similar spin-polarization induced by C- (B- and N-) anion doping also appears in rutile TiO$_2$, see Figure \ref{f21}.

It is noted that the magnetic property of C- (N- and B-) doped ZnO and ZnS has also been well studied from first-principles calculations.\cite{Pan_PRL_2007, Yang_PRB_2010, Shen_PRB_2008, Peng_2009_APL, Long_2009_PRB, Fan_2009_APL, Wang_2009_PLA, Chen_2011_PLA} One C (N and B) atom at an O site in ZnO produces a spin magnetic moment of 2.0 (1.0 and 3.0) {$\mu$}$_B$, respectively, and thus it is proposed that the magnetic moment per dopant (in {$\mu$}$_B$) is equal to the difference of the atomic numbers between the dopant and host anion in AlN- and ZnO-based semiconductors.\cite{Peng_2009_APL} This is reasonable for AlN- and ZnO-based semiconductors. For X-anion (X=C, N, and B) doped AlN- and ZnO-based semiconductor, the local XZn$_4$ structure has a $\emph{T$_d$}$ symmetry, in which the 2$\emph{p}$ orbitals of dopant are 3-fold degenerated, and thus C$^{2-}$ ($\emph{s$^2$p$^4$}$), N$^{2-}$ ($\emph{s$^2$p$^5$}$), and B$^{2-}$($\emph{s$^2$p$^3$}$) produces a spin magnetic moment of 2.0, 1.0, and 3.0 {$\mu$}$_B$, respectively. For X-anion doped AlN-based semiconductor, the doped C, N, and B exist as C$^{3-}$($\emph{s$^2$p$^5$}$), N$^{3-}$($\emph{s$^2$p$^6$}$), and B$^{3-}$($\emph{s$^2$p$^4$}$) ions, and the distribution of the 2$\emph{p}$ electrons on the 3-fold $\emph{p}$ orbitals produces a spin magnetic moment of 1.0, 0, and 2.0 {$\mu$}$_B$, respectively. In contrast, the local CTi$_3$ structure of C-doped TiO$_2$ has a $\emph{C$_{2v}$}$ symmetry, and the 3-fold degenerate $\emph{p}$ orbitals of B atom are split into three nondegenerate energy levels. In B-doped anatase TiO$_2$, the three $\emph{p}$ electrons of B$^{2-}$ ($\emph{s$^2$p$^3$}$) preferentially occupy the up-spin and down-spin 2$\emph{p$_x$}$ and up-spin 2$\emph{p$_z$}$ orbitals , thus generating a magnetic moment of 1.0 {$\mu$}$_B$. Therefore, the symmetry of the local structure is a critical factor to analyze spin magnetic moment besides the difference of the fundamental property between the dopant and host anion.

\subsection{Magnetic Coupling}
\begin{figure*}
\center
\includegraphics[scale=0.40]{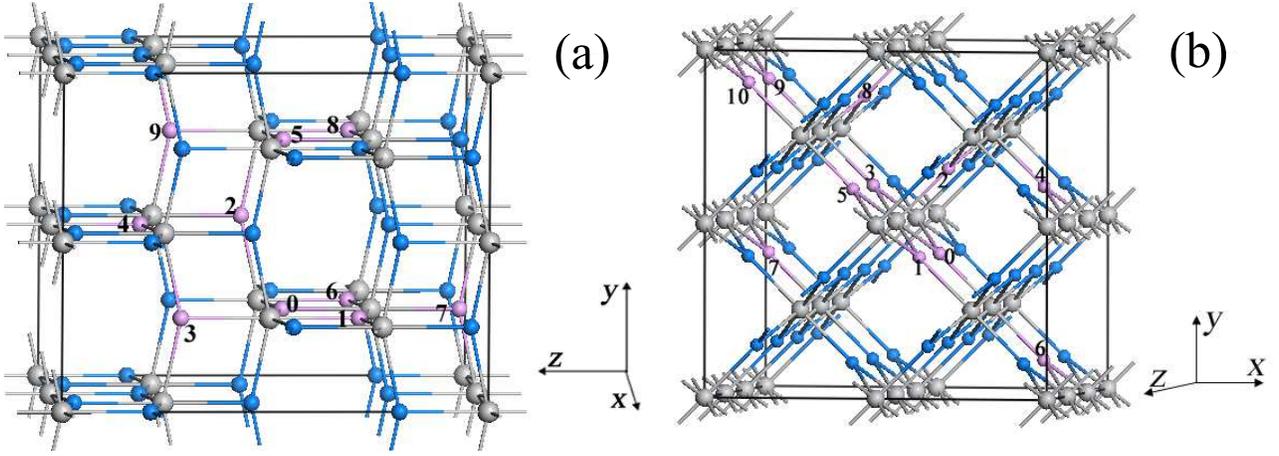}
\caption{(Color online) (a) 48-atom 2$\times$2$\times$1 supercell of anatase phase employed to define C (N)-doped TiO$_2$ structure. (b) 72-atom 2$\times$2$\times$3 supercell of rutile phase employed to define C (N)-doped TiO$_2$ structure. The larger and small spheres represent the Ti and O atoms, respectively. The O atoms labeled by 0-9 in anatase TiO$_2$ and labeled by 0-10 in rutile TiO$_2$ are the sites to be replaced with C (N) atoms.
Reprinted with permission from Yang, K.; Dai, Y.; Huang, B.; Whangbo, M.-H.
\href{http://dx.doi.org/10.1063/1.2996024} {\emph{Appl. Phys. Lett.} \textbf{2008}, 93, 132507.} Copyright (2008) American Institute of Physics and Yang, K.; Dai, Y.; Huang, B.; Whangbo, M.-H.
\href{http://dx.doi.org/10.1016/j.cplett.2009.09.050} {\emph{Chem. Phys. Lett.} \textbf{2009}, 481, 99.} Copyright (2009) Elsevier.
}\label{f22}
\end{figure*}

Whether a stable RTFM can be formed is determined by the magnetic coupling strength between the magnetic moments of dopants, which is reflected by the energy difference between the ferromagnetic (FM) and antiferromagnetic (AFM) states. To explore the possibility of ferromagnetism in C- (N- and B-) doped TiO$_2$, ten different arrangements of TiO$_2$ with two dopants (48-atom 2$\times$2$\times$1 supercell for the anatase phase and 72-atom 2$\times$2$\times$3 supercell for the rutile phase) were modeled to examine the magnetic coupling strength (see Figure \ref{f22}). For doped anatase TiO$_2$, ten arrangements of two dopants are modeled by replacing two O atoms using two C (N) atoms at positions (0, 1), (0, 2), (0, 3), (0, 4), (0, 5), (0, 7), (0, 8), (0, 9), (1, 6), and (3, 8). For doped rutile TiO$_2$, the ten arrangements are modeled by replacing two O atoms using two C (N) atoms at positions (0, 1), (0, 2), (0, 3), (0, 4), (0, 5), (0, 6), (0, 7), (0, 8), (0, 9), and (0, 10).

\begin{figure}
\center
\includegraphics[scale=0.36]{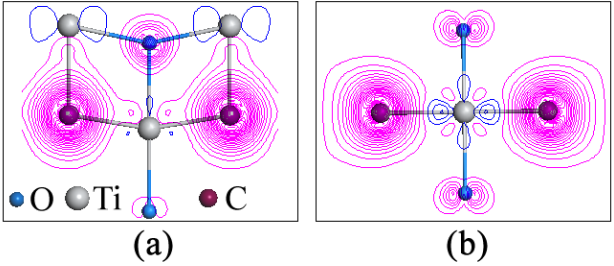} 
\caption{(Color online) Spin density distributions under FM alignment around the C atoms for configuration (1, 6) of anatase TiO$_2$, and configuration (0, 6) of rutile TiO$_2$. The pink and blue contour lines indicate the up-spin and down-spin density, respectively.}
\label{f23}
\end{figure}

For C-doped anatase and rutile TiO$_2$, both the configurations of (0, 1), (0, 2) and (0, 3) are nonmagnetic, in which the C$\cdots$C distance are shorter or close to a typical C=C double bond, see Table \ref{t2} for anatase TiO$_2$ and Table \ref{t3} for rutile TiO$_2$. These nonmagnetic configurations are considerably more stable than other configurations. This indicates that the C dopants have a tendency to form a C cluster through direct CC bonding interaction. In particular, the (1, 6) configuration of anatase TiO$_2$ and (0, 6) configuration of rutile TiO$_2$ form a stable FM coupling. To understand their FM coupling mechanism, the spin density distributions around the two C$^{2-}$ ions of the two configurations are presented in Figure \ref{f23}. It shows that the two C$^{2-}$ ions of (1, 6) configuration of anatase TiO$_2$ and the (0, 6) configuration of rutile TiO$_2$ both have a strong spin-polarization, and the spin-polarization extends to the second-nearest-neighbor O ions. The magnetic orbitals of the two C$^{2-}$ ions can overlap around the common second-nearest-neighbor O ions, so that the overlapped spin density might lead to the FM coupling between the magnetic moments of the two C$^{2-}$ ions.\cite{Hay_1975_JACS} The theoretical prediction was later confirmed by the experiments in which RTFM was found in C-doped TiO$_2$.\cite{Ye_PLA_2009, Wen_IEEE.Trans.Magn_2009} In contrast, a stable AFM or paramagnetic (PM) state was formed in the other structures, in which their spin density plots show that the two C dopants do not have common spin-polarized O ions. Therefore, the AFM might result from the direct magnetic orbital interaction between the two C$^{2-}$ ions. When the distance between the two C sites increases, a PM state is formed in which the magnetic interaction between the magnetic orbitals of the two C ions becomes negligible.

\begin{table}
\centering
\caption{Values of the C$\cdots$C distance, the relative energies $\Delta E$ (eV) and $E_{mag}$ (meV), and the magnetic moment ($M$) ($\mu_B$) on each C atoms calculated for the (i, j) structure of the two-C-atom doped anatase TiO$_2$. For each (i, j) structure, the energy of the lower-energy state (either FM or AFM) is given with respect to that of the (0, 1) structure, i.e., $\Delta E$=E(i, j)-E(0, 1). For each (i, j) structure, the energy of the FM state is given with respect to that of the AFM state, i.e., \emph{E$_{mag}$}=\emph{E$_{FM}$}-\emph{E$_{AFM}$} (meV).
Reprinted with permission from Yang, K.; Dai, Y.; Huang, B.; Whangbo, M.-H.
\href{http://dx.doi.org/10.1063/1.2996024}
{\emph{Appl. Phys. Lett.} \textbf{2008}, 93, 132507.} Copyright (2008) American Institute of Physics.
}
\begin{tabular} {ccccc}
\hline
(i, j)	&	C$\cdots$C	&	$\Delta E$ (eV) &	$E_{mag}$ (meV)	 &	$M$ ($\mu_B$)	 \\
\hline
(0, 1)	&	1.339	&	0	&	~0	&	0	\\
(0, 2)	&	1.294	&	0.27	&	~0	&	0	\\
(0, 3)	&	1.274	&	1.16	&	~0	&	0	\\
(0, 4)	&	3.027	&	6.43	&	106	&	0.67	\\
(0, 5)	&	3.776	&	6.55	&	102	&	0.66	\\
(1, 6)	&	3.776	&	6.38	&	-251	&	0.77	\\
(0, 7)	&	4.126	&	6.17	&	75	&	0.53	\\
(0, 8)	&	4.644	&	6.53	&	16	&	0.69	\\
(0, 9)	&	4.973	&	6.39	&	48	&	0.67	\\
(3, 8)	&	6.825	&	6.63	&	5	&	0.69	\\
\hline
\end{tabular}
\label{t2}
\end{table}

\begin{table}
\centering
\caption{Calculated results for C-doped rutile TiO$_2$.}
\begin{tabular} {ccccc}
\hline
(i, j)	&	C$\cdots$C	&	$\Delta E$ (eV) &	$E_{mag}$ (meV)	 &	$M$ ($\mu_B$)	 \\
\hline
(0, 1)	&	1.314	&	0	&	0	&	0	\\
(0, 2)	&	1.282	&	0.18	&	0	&	0	\\
(0, 3)	&	1.368	&	2.16	&	0	&	0	\\
(0, 4)	&	3.152	&	5.28	&	12	&	0.68	\\
(0, 5)	&	4.11	&	4.88	&	103	&	0.52	\\
(0, 6)	&	4.036	&	5.56	&	-192	&	0.73	\\
(0, 7)	&	4.643	&	5.37	&	9	&	0.68	\\
(0, 8)	&	5.019	&	5.38	&	8	&	0.69	\\
(0, 9)	&	6.605	&	5.57	&	8	&	0.68	\\
(0, 10)	&	7.358	&	5.36	&	-5	&	0.69	\\
\hline
\end{tabular}
\label{t3}
\end{table}

For N-doped anatase and rutile TiO$_2$, a systemic study of the magnetic coupling interaction between two N dopants was also performed.\cite{Yang_CPL_2009, Bai_2008_SSC, Tao_APL_2009} Two important conclusions could be obtained as below\cite{Yang_CPL_2009}:

(a)	A substantial FM or AFM coupling occurs when two N$^{2-}$ ions are coordinated to a common Ti ion. A stable AFM coupling occurs when the $\angle$N-Ti-N angle is small (less than $\sim$99$^0$), and a stable FM coupling occurs when the $\angle$N-Ti-N angle is large (greater than $\sim$102$^0$).

(b)	The $\emph{$\Delta$E}$ values for the various configurations in N-doped anatase and rutile TiO$_2$ are nearly uniform. This indicates that the N dopants do not have a tendency to form a cluster, unlike the case of C-doped TiO$_2$.\cite{Yang_APL_2008}

\begin{figure}
\center
\includegraphics[scale=0.5]{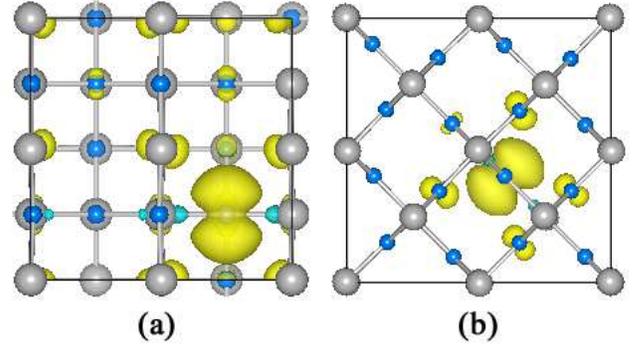} 
\caption{(Color online) Spin density distribution around the N-dopant calculated for one-N-atom doped (a) anatase and (b) rutile TiO$_2$.
Reprinted with permission from Yang, K.; Dai, Y.; Huang, B.; Whangbo, M.-H.
\href{http://dx.doi.org/10.1016/j.cplett.2009.09.050} {\emph{Chem. Phys. Lett.} \textbf{2009}, 481, 99.} Copyright (2009) Elsevier.
}\label{f24}
\end{figure}

The three-dimensional (3D) spin density distribution plots of one-N-atom doped anatase and rutile TiO$_2$ are presented in Figure \ref{f24}. The magnetic orbital around the N dopant has a dumbbell-shaped distribution, the axis of which is perpendicular to the NTi$_3$ plane, and the spin density mostly originates from the N 2$\emph{p}$ orbital. A detailed analysis of the 3D spin density distribution for the two-N-atom doped anatase and rutile TiO$_2$ shows that the magnetic orbital interaction between the two N sites can be divided into $\delta$-type, $\pi$-type and orthogonal ($\bot$) type. The strength and type of the orbital interaction could be responsible for the FM/AFM coupling mechanism. When the $\angle$N-Ti-N angle is small (less than $\sim$99$^0$), a substantial through-space $\delta$- or $\pi$-type magnetic orbital interaction between the two N-dopants leads to an AFM coupling. A direct through-space $\delta$-type overlap also takes place in the (0, 5) configuration of anatase TiO$_2$ and the (0, 7) configuration of rutile TiO$_2$, in which an AFM coupling is formed even though the two N dopants are not coordinated the same Ti ion. When the $\angle$N-Ti-N angle is large, the through-space $\delta$- or $\pi$-type magnetic orbital interaction becomes negligible, and in this case, the magnetic coupling interaction gives rise to FM coupling and its ground FM state is more stable than AFM state about from 10 to 39 meV. This is consistent with the experimentally observed ferromagnetism in N-doped rutile TiO$_2$.\cite{Cruz_JPCM_2009, Bao_2011_JAP} It is worth noting that the effective magnetic coupling interaction only occurs between the nearest neighbor N dopants. This means that there exists a percolation threshold of nitrogen-concentration to produce the ferromagnetism. Only if the nitrogen-concentration reaches a minimal value of about 14\%, can ferromagnetism be observed.\cite{Yang_CPL_2009} A similar threshold of the nitrogen-concentration to produce ferromagnetism also occurs in N-doped SrTiO$_3$ and BaTiO$_3$.\cite{Yang_2011_APL}

For B-anion doped TiO$_2$, a detailed discussion of the relative stability and magnetic coupling characteristics of two-B-doped anatase TiO$_2$ has been done.\cite{Yang_2010_JPCC_B} With respect to the structural configuration in Figure \ref{f22}, (0, 1), (0, 2), (0, 3) and (0, 4) structures of two-B-atom doped anatase TiO$_2$ are nonmagnetic, the B$\cdots$B distances of which are much closer to the bond length of boron cluster (1.48$\sim$1.7{\AA}),\cite{Reis_2000_IJQC}  and they are relatively more stable than other magnetic structures [(0, 5), (1, 6) and (3, 8)]. This indicates that the B atoms prefer to form a cluster through direct B-B bonding interactions, similar to the case of C-doped ZnO and TiO$_2$.\cite{Pan_PRL_2007, Yang_APL_2008} In (0, 8) and (0, 9) structures, the B dopants are displaced from the oxygen sites upon structural relaxation, forming interstitial B-doped structure. Total energies of these two structures are much close to those of (0, 1), (0, 2), (0, 3) and (0, 4) configurations, thus indicating that the interstitial B-doped structure is relatively easy to form.\cite{Geng_2006_JPCM} It is also noted that the magnetic coupling interaction between the magnetic orbitals of two B$^{2-}$ is much weaker than that of C (N)-doped TiO$_2$.\cite{Yang_APL_2008, Yang_CPL_2009, Tao_APL_2009, Bai_2008_SSC} In spite of their magnetic coupling types, the maximum total energy difference between the FM and AFM state is about 27 meV, and it becomes negligible when the B$\cdots$B distance exceeds about 7{\AA}. Therefore, B-anion doping is not an effective way to produce $\emph{d$^0$}$ ferromagnetism in TiO$_2$.

\section{Conclusion and Outlook}
A deep theoretical understanding on the structural stability, electronic structure, optical and magnetic properties of nonmetal-doped TiO$_2$ is of great importance to develop its photocatalytic and spintronic applications, which can be well established from first-principles calculations. First-principles theoretical calculation of doped TiO$_2$ is still an ongoing subject. Herein, we would like to point out future theoretical direction from the viewpoint of first-principles calculations. One is the nonmetal or metal doping influence on the anatase to rutile phase transition, and a clear theoretical understanding on its mechanism will be interesting and useful to optimize the performance of TiO$_2$ in photocatalytic and other applications. Another one is to explore an effective physical or chemical parameter to characterize the photocatalytic performance of TiO$_2$, which should be directly or indirectly obtained from first-principles calculations. If so, some approaches can be taken to maximize the photocatalytic activity of TiO$_2$.

\begin{acknowledgements}
This work is supported by the National Basic Research Program of China (973 program, 2007CB613302), National Science foundation of China under Grant 11174180 and 20973102, and the Natural Science Foundation of Shandong Province under Grant number ZR2011AM009.
\end{acknowledgements}


\end{document}